\documentclass[preprint,3p,times,onecolumn]{elsarticle} 
\usepackage{graphicx}
\usepackage{epsfig}
\usepackage{amsmath}
\usepackage{amsfonts}
\usepackage{amssymb}
\usepackage{dcolumn}
\usepackage{bm}
\usepackage{dcolumn}
\usepackage{color}
\usepackage{verbatim}
\usepackage{paralist}
\usepackage{amssymb}
\usepackage{graphicx}
\usepackage{extarrows}
\usepackage{url}
\usepackage{tabularx}
\usepackage{lineno}
\usepackage{etoolbox}
\usepackage[section]{placeins}
\usepackage[usestackEOL]{stackengine}

\usepackage{color}

\def\lapprox{\mathrel{\mathop  {\hbox{\lower0.5ex\hbox{$\sim$}
\kern-1.1em\lower-0.7ex\hbox{$<$}}}}}
\def\gapprox{\mathrel{\mathop  {\hbox{\lower0.5ex\hbox{$\sim$}
\kern-1.1em\lower-0.7ex\hbox{$>$}}}}}

\usepackage{etoolbox}

\begin{document}

\begin{frontmatter}

\title {Search for low-energy neutrinos from astrophysical sources with Borexino }

\author[Munchen]{M.~Agostini}
\author[Munchen]{K.~Altenm\"{u}ller}
\author[Munchen]{S.~Appel}
\author[Kurchatov]{V.~Atroshchenko}
\author[Juelich]{Z.~Bagdasarian}
\author[Milano]{D.~Basilico}
\author[Milano]{G.~Bellini}
\author[PrincetonChemEng]{J.~Benziger}
\author[Hamburg]{D.~Bick}
\author[LNGS]{G.~Bonfini}

% ...

\fntext[Roma]{Present address: Dipartimento di Fisica, Sapienza Universit\`a di Roma e INFN, 00185 Roma, Italy}
\fntext[Napoli]{Present address: Dipartimento di Fisica, Universit\`a degli Studi Federico II e INFN, 80126 Napoli, Italy}
\fntext[Madrid]{Present address: Universidad AutÛnoma de Madrid, Ciudad Universitaria de Cantoblanco, 28049 Madrid, Spain}
\fntext[Fermi]{Present address: Fermilab National Accelerato Laboratory (FNAL), Batavia, IL 60510, USA}
\fntext[LNGSG]{Present address: INFN Laboratori Nazionali del Gran Sasso, 67010 Assergi (AQ), Italy}
\fntext[KFKI]{Also at: MTA-Wigner Research Centre for Physics, Department of Space Physics and Space Technology, Konkoly-Thege MiklÛs ˙t 29-33, 1121 Budapest, Hungary}
\tnotetext[fn1]{Deceased in August 2019}
  
\address{\bf{The Borexino Collaboration}}

% ...
\address[Kurchatov]{National Research Centre Kurchatov Institute, 123182 Moscow, Russia}
% ...
\address[LNGS]{INFN Laboratori Nazionali del Gran Sasso, 67010 Assergi (AQ), Italy}
\address[Milano]{Dipartimento di Fisica, Universit\`a degli Studi e INFN, 20133 Milano, Italy}
% ...
\address[PrincetonChemEng]{Chemical Engineering Department, Princeton University, Princeton, NJ 08544, USA}
% ...
\address[Munchen]{Physik-Department and Excellence Cluster Universe, Technische Universit\"at  M\"unchen, 85748 Garching, Germany}
% ...
\address[Juelich]{Institut f\"ur Kernphysik, Forschungszentrum J\"ulich, 52425 J\"ulich, Germany}
% ...
\address[Hamburg]{Institut f\"ur Experimentalphysik, Universit\"at Hamburg, 22761 Hamburg, Germany}

\begin{abstract}

We report on searches for neutrinos and antineutrinos from astrophysical sources performed with the Borexino detector at the Laboratori Nazionali del Gran Sasso in Italy. Electron antineutrinos ($\bar{\nu}_e$) are detected in an organic liquid scintillator through the inverse $\beta$-decay reaction. In the present work we set model-independent upper limits in the energy range 1.8-16.8 MeV on neutrino fluxes from unknown sources that improve our previous results, on average, by a factor 2.5. Using the same data set, we first obtain experimental constraints on the diffuse supernova $\bar{\nu}_e$ fluxes in the previously unexplored region below 8 MeV.
A search for $\bar{\nu}_e$  in the solar neutrino flux is also presented: the presence of $\bar{\nu}_e$ would be a manifestation of a non-zero anomalous magnetic moment of the neutrino, making possible its conversion to antineutrinos in the strong magnetic field of the Sun. We obtain a limit for a solar $\bar{\nu}_e$ flux of 384 cm$^{-2}$s$^{-1}$ (90\% \text{C.L.}), assuming an undistorted  solar $^{8}$B neutrinos energy spectrum, that corresponds to a transition probability  $p_{ \nu_e \rightarrow \bar\nu_{e}}<$ 7.2$\times$10$^{-5}$ (90\% \text{C.L.}) for E$_{\bar {\nu}_e}$ $>$ 1.8 MeV. 
At lower energies, by investigating the spectral shape of elastic scattering events, we obtain a new limit on solar $^{7}$Be-$\nu_e$ conversion  into $\bar{\nu}_e$ of $p_{ \nu_e \rightarrow \bar \nu_{e}}<$ 0.14 (90\% \text{C.L.}) at 0.862 keV. Last, we investigate solar flares as possible neutrino sources and obtain the strongest up-to-date limits on the fluence of neutrinos of all flavor neutrino below 3--7\,MeV. Assuming the neutrino flux to be proportional to the flare's intensity, we exclude an intense solar flare as the cause of the observed excess of events in run 117 of the Cl-Ar Homestake experiment.

\end{abstract}


\end{frontmatter}
\twocolumn
\sloppy %

\section{Introduction}
\label{sec:intro}
Astrophysical neutrinos cover at least 18 orders of magnitude in energy, starting from meV (relic neutrinos) till PeV, the highest energy neutrinos ever detected as of today.
Alongside the detection of gravitational waves \cite{Abbott 2016}, an event that has opened a new era of gravitational astronomy, collecting more data on astrophysical neutrinos and discovering their possible new sources will affect the very foundations of our understanding of the Universe.
Neutrino detectors indeed start playing a substantial role in multi-messenger astronomy \cite{neutrino_from_blazar}.

So far, neutrino astronomy has accumulated a wide range of experimental achievements, including the detection of neutrinos from supernova SN1987A \cite{Hirata 1987, Bionta 1987, Alexeyev 1988, Aglietta 1987}, the detection of extragalactic neutrinos with energies up to 2\,PeV \cite{Aartsen 2013}, and the precision spectroscopy of neutrinos from the Sun \cite{Agostini:2018uly}. 
For some neutrino sources the accumulation of statistics is ongoing, while others do not have experimental confirmation yet. 

Borexino detector has proven its potential in the various fields of experimental neutrino astronomy.
Among the recent achievements are the precise spectral measurements of neutrinos originating in different nuclear fusion reactions of the \textit{pp} chain in the Sun \cite{Agostini:2018uly}, the best upper limits on the neutrino and antineutrino fluences of all flavors from gamma-ray bursts in the energy range $E_{\nu} <$ 7\,MeV \cite{BxGRB}, and the best upper limits on all flavor neutrino fluence associated with gravitational wave events within 0.5--5.0\,MeV energy range \cite{Borexino GW}.
This paper is aimed to explore the possible existence of tiny antineutrino fluxes associated to extraterrestrial sources, as well as to search for possible neutrino signals time-correlated with solar flares.

A solar flare is a sudden flash of increased brightness on the Sun: powerful flares are often accompanied by a coronal mass ejection. 
If the ejection occurs in the direction of the Earth, related particles can penetrate into the upper atmosphere, cause bright auroras, and even disrupt long range radio communication. 
Neutrinos could be emitted in correlation with solar flares: the protons accelerated in the regions of magnetic reconnection occurred during the flare, may produce pions through a number of nuclear collisions in plasma. Electron and muon neutrinos in the MeV-GeV range may originate in the sequential decays of pions and muons and be ejected from the Sun.
The detection of neutrinos in the solar flares would provide a deeper understanding of nuclear processes in the solar atmosphere.
The experimental studies began  after the attempt to attribute an excess of neutrino events observed in several runs taken  by the Homestake experiment \cite{SFlares Homestake-1, SFlares Homestake-2} to the solar flares.
Thereafter, the search for solar flare neutrinos in MeV energy range was pursued by Kamiokande \cite{SFlares Kamioka 1, SFlares Kamioka 2}, LSD \cite{SFlares LSD}, and SNO \cite{SFlares SNO}, and limits on neutrino fluence were  set. 

Among the possible extraterrestrial sources of antineutrinos are the supernovae explosions and the conversion of solar $\nu_e \rightarrow \bar{\nu}_e$ in the strong solar magnetic field in the case of an anomalous neutrino magnetic moment.

The Diffuse Supernova Neutrino Background 
(DSNB, sometimes referred to as supernova relic neutrinos) is formed by the whole of the star collapsing during the evolution of the Universe and consists of neutrinos and antineutrinos of all flavors.
The study of DSNB energy spectra allows us to address important issues of neutrino physics, astrophysics, and cosmology.

The observed DSNB flux spectrum is given by \cite{Ando 2004}:
\begin{equation} \frac{d\phi_\nu}{dE_\nu} = \frac{c}{H_{0}} \int_{0}^{z_{max}} \frac{dN_\nu(E_{\nu}')}{dE_{\nu}'} \frac{R_{SN}(z) dz} {\sqrt{\Omega_{m}(1+z)^{3}+\Omega_{\Lambda}}},
\label{eq:fluxdsnb}
\end{equation} 
 where $c$ is the speed of light, $H_{0}$ is Hubble constant, $z$ is the red shift, $\frac{dN_\nu(E_{\nu}')}{dE_{\nu}'}$ is the neutrino emission spectrum for individual supernova, $R_{SN}(z)$ is the supernova rate at the distance $z$ to the observer, $\Omega_{m}$ and $\Omega_{\Lambda}$ are the relative densities of matter and dark energy in the Universe, respectively.
 The spectrum is sensitive to particular cosmological model through $\Omega_{m}$ and $\Omega_{\Lambda}$ and reflects the expansion of the Universe through the dependence on $H_{0}$.
Other impacts of DSNB studies are neutrino properties, due to the dependence of $\frac{dN_\nu(E_{\nu}')}{dE_{\nu}'}$ on neutrino mass hierarchy, on neutrino magnetic moment, and on the still not-excluded non-standard interactions of neutrinos.
In recent years, several experiments searched for DSNB. 
The KamLAND collaboration set an upper limit for the diffuse supernova $\bar{\nu}_e$ flux of 139\,cm$^{-2}$s$^{-1}$ (90\% \text{C.L.}) in the energy range 8.3--31.8~MeV \cite{KamLAND 2012}.
The Super-Kamiokande set an upper bound of 2.9 $\bar{\nu}_e$ cm$^{-2}$s$^{-1}$ (90\% \text{C.L.}) in the energy region $E_{\bar \nu_{e}} >$ 17.3\,MeV \cite{Super-K 2012}. 
For ${\nu}_e$ the upper limit of 70\,cm$^{-2}$s$^{-1}$ (90\% \text{C.L.}) was inferred by Sudbury Neutrino Observatory in the energy range 22.9--36.9~MeV \cite{SNO 2006}.

Apart from DSNB, we performed a search for antineutrinos in the solar neutrino flux. The presence of $\bar{\nu}_e$ would be a manifestation of the anomalous magnetic moment of neutrino, making possible its conversion to antineutrino in the magnetic field of the Sun due to combined effects of the spin-flavor precession (SFP) and neutrino oscillations.
Borexino collaboration has published the results of a model-independent study of solar and other unknown antineutrino fluxes in 2011 \cite{BxAntinu}.
Based on five-fold increase in statistics, this paper presents an update on model-independent search and includes model-dependent limits on DSNB antineutrinos and solar neutrino conversion.

The paper is structured as follows. 
After an overview of the  Borexino detector layout (Section~\ref{sec:detector}),  the investigation for extraterrestrial $\bar{\nu}_e$ fluxes is detailed in Section~\ref{sec:dsnb}: first the events selection cuts and the background sources are described, then we present the results divided into three main topics: (1) model-independent $\bar{\nu}_e$ analysis, (2)  search for DSNB $\bar{\nu}_e$ and (3) limits on the solar $\nu_e$ into $\bar{\nu}_e$ conversion probability.  
On this last topic we present the limits obtained not only by looking for the inverse $\beta$-decay signals but also the ${\bar \nu_{e}}$ elastic scattering process,  below  the inverse $\beta$-decay  threshold. 

At last, Section~\ref{sec:flares} is devoted to the search for neutrino signals associated to solar flares.

\section{The Borexino experiment}
\label{sec:detector}
Borexino is an unsegmented liquid scintillation detector built for the spectral measurement of low--energy solar neutrinos installed in the underground hall C of the Laboratori Nazionali del Gran Sasso (LNGS) in Italy.
The target mass is made of 278\,tons of ultra--pure liquid scintillator (pseudocumene (PC) doped with 1.5~g/l of diphenyloxazole) and enclosed within a spherical nylon inner vessel (IV) with a radius of 4.25\,m. 
The detector core is shielded from external radiation by 890 tons of buffer liquid, a solution of PC and 2-5~g/l of the light quencher dimethylphthalate.
The buffer is divided in two volumes by the second nylon vessel with a 5.75~m radius, preventing inward radon diffusion and transfer by convection.
All these volumes are contained in a 13.7~m diameter stainless steel sphere (SSS) on which are mounted 2212~8'' photomultiplier tubes (PMT) detecting the scintillation light, forming the so--called Inner Detector (ID).  
The choice of a liquid scintillator as target  is especially important to observe the low energy neutrino events: the high light yield typical of Borexino scintillator ($\sim$10$^{4}$\,photons/MeV)  makes it possible to get a good energy resolution and to set  a very low energy threshold (50\,keV). The high transparency (the attenuation length is close to 10 m at 430\,nm) and  the fast time response (a few ns) allow for  an event position reconstruction  and a good pulse shape discrimination (PID) between alpha and beta/gamma decays. 

The SSS is immersed in a water tank of 9\,m radius and 16.9\,m height (Outer Detector, OD), filled with ultra--high purity water and instrumented with 208 PMTs serving as a \v{C}erenkov active muon veto.
The water contained in the water tank ($\sim$2\,m, at least, around the SSS in all directions; 2400\,m$^{3}$ in total) also provides good shielding with respect to gammas and neutrons emitted by the rocks and by the surrounding laboratory environment. 
A more detailed description of the Borexino detector can be found in~\cite{BXdetector}. 
Several calibration campaigns with radioactive sources~\cite{BXcalib} allowed a decrease in the systematic errors of the measurements and to optimize Geant4 based Monte Carlo (MC) simulation code \cite{g4bx}.
The present analysis is based on two semi-independent data acquisition systems: the primary Borexino readout optimized for solar neutrino physics up to a few MeV and a fast waveform
digitizer system tuned for events above 1\,MeV~\cite{BXdetector}. 
The primary electronics of Borexino, in which all 2212 channels are read individually, is optimized for energies up to few MeV.
The energy ($E$) of each event is reconstructed using the total amount of light registered by all PMTs of the detector, measured as a charge ($N_{p.e.}$) in photoelectrons (p.e.), and corrected with a position and time dependent light collection function $E$=$f(N_{p.e.}; x,y,z,t)$. This function was constructed using MC simulations, checked against calibration data from radioactive sources deployed in different positions inside the detector~\cite{BXcalib,g4bx}. The reason for the time dependence is that the number of working PMTs has been declining with time, with a reduction of  $\sim$35\% in 10 years.
A typical energy deposit of 1\,MeV at the center of the detector produces a signal of about 500 photoelectrons (normalized to 2000 PMTs), resulting in an energy resolution of $\sim 5\%$~$/ \sqrt {E(\rm{MeV})}$.

For higher energies, up to $\sim$50 MeV, a system was developed consisting of 96 fast waveform digitizers (CAEN v896, 8 bit, thereafter FADC - Flash ADC), each of them reading-in the signal summed from up to 24 PMTs, with the sampling rate of 400\,MHz. The FADC DAQ energy threshold is $\sim$1\,MeV.
Starting from December 2009, it acquires data in a new hardware configuration, having a separate trigger.
FADC energy scale is calibrated using the 2.22\,MeV gamma peak originating
from cosmogenic neutron captures on protons, as well as by fitting the
spectrum of cosmogenic $^{12}$B and Michel electrons from muons decaying
inside the detector.
The energy resolution of the FADC system was found to be $\sim 10\%$~$/  \sqrt {E(\rm{MeV})}$.
The two DAQ systems are synchronized and merged offline with a special software utility based on a GPS time of each trigger.
Advantages of each system, such as higher energy resolution in the primary DAQ or advanced software algorithms for muon and electronics-noise tagging for energies above 1\,MeV in the FADCs, were exploited in different ways depending on the analysis, as specified below.

\subsection{Muon tagging}
\label{subsec:veto}
Despite the 3800 m.w.e. of the rock overburden, the muon flux at LNGS is still significant (1.2 muon/m$^{2}$/h). The number of muons crossing the whole detector is $\sim$8600/day~\cite{muons,Cosmogenic} and half of them, on average, deposit some energy in the Inner Detector. 
The muons in Borexino are identified both by the signal released in the outer detector (Outer Detector Flag) and/or by the space and time distributions of the emitted light in the inner detector (Inner Detector Flag). The latter distributions are quite different in respect to the one induced in point-like interactions, e.g., in the low energy neutrino ones.  Due to the OD veto and the pulse shape analysis of ID tracks, the muon background can be reduced by a factor of 10$^{5}$.
The standard Borexino muon identification~\cite{muons,Cosmogenic}, including both OD veto and ID pulse-shape muon tagging, is prone to mistakenly tag point-like events above $\sim$15\,MeV as muons.
In this work, we have optimized the muon cuts for energies up to 20\,MeV that covers all the present analyses,  by modifying the muon tagging based on the ID pulse-shape.
In particular, the threshold values for the variables describing the mean duration and the peak position of the light pulse have been accurately tuned with the  Borexino simulation code, and further criteria have been added based on the anisotropy variable $S_p$~\cite{LongPaper} as well as on the reconstructed radial position.
In addition, muon tagging based on the FADC waveform analysis was developed and tested against the sample of muons detected by the OD: the overefficiency to tag point-like events at energies up to few tens of MeV as muons was cross-checked on a sample of Michel electrons.
In all analyses, events associated with muons have been removed. 

Muons interacting with carbon nuclei in the scintillator or with the surroundings materials can produce neutrons, high energy gammas and a variety of radioactive isotopes.
To strongly suppress this potential background, the events following every tagged (and removed) muon are excluded from the data sample: a 2\,ms veto is applied after muons crossing only the OD to remove  penetrating neutrons. In addition, a veto window after muons crossing the ID is applied, to suppress cosmogenic isotopes. A different time length for this window was chosen, depending on the analysis, from 0.3 to 2\,s as detailed below.

\subsection{Fiducial volume definition}
\label{subsec:fv}

The shape of the IV enclosing the Borexino scintillator deviates from a sphere and is changing in time, because of buoyancy effects caused by the different densities of the 
buffer and  scintillator liquids, as well as a consequence of a small IV leak that started approximately in April 2008.
To track the evolution of the vessel shape is thus crucial.  
The vessel is more contaminated with respect to the scintillator:  the events originated by the radioactive contaminations of the nylon, in particular $^{210}$Bi, $^{14}$C and $^{208}$Tl, are selected to reconstruct the time-dependent IV shape.
The procedure was cross-checked and calibrated over several ID pictures taken throughout the years with an internal CCD camera system: the precision of the method is found to be of the order of $\pm$5\,cm.
For each analysis a {\it Dynamical Fiducial Volume} (DFV) is defined by considering the time-dependent vessel shape. Antineutrinos' study usually allows for the choice of a shallow FV cut (25 cm from the vessel) while the study of neutrino-scattering events at low energies ($<$3\,MeV) requires a stronger suppression of the external gamma background and, therefore, a smaller and innermost FV (75\,cm from the vessel).

\section{The search for extraterrestrial ${\bar \nu_{e}}$ fluxes}
\label{sec:dsnb}
The possible signal due to tiny, still undisclosed, extraterrestrial $\bar{\nu}_e$ fluxes, such as supernovae relic neutrinos, can be put in evidence as an excess of events with respect to the backgrounds and the known sources of $\bar{\nu}_e$.
A new study is presented here that benefits from the exceptional radiopurity of the Borexino detector and extends over a  wide energy range, from 1.8 to 16.8\,MeV.
In Sec.~\ref{subsec:dsnb_antinu_data} we describe the data set and the events selection cuts, in Sec.~\ref{subsec:dsnb_antinu_backg} the sources of backgrounds, then the selected events are presented (Sec. \ref{subsec:events}) followed by the model independent limits for $\bar{\nu}_e$ fluxes (Sec.~\ref{subsec:dsnb_antinu_womod}) and by limits that assume the $\bar{\nu}_e$ spectral shape predicted by two different recent models of DSNB (Sec. \ref{subsec:dsnb_antinu_wmod}). Finally in Sec. \ref{subsec:sfp} we present the study of solar ${ \nu_{e}}$  conversion into $\bar{\nu}_e$ together with the corresponding upper limits for the neutrino magnetic moment.

\subsection{Dataset and events selection} 
\label{subsec:dsnb_antinu_data}

Electron antineutrinos are detected in Borexino through the reaction of Inverse Beta Decay of the free proton (IBD):
\begin{equation}
\bar{\nu}_e+p\to n+e^{+}
\end{equation}
The energy threshold of this reaction is $E_{\bar {\nu}_e}$ = 1.8\,MeV. 
The positron deposits its kinetic energy and annihilates almost immediately, inducing a prompt signal. The light produced in the scattering of the positron is intrinsically indistinguishable from the light produced by two annihilation gammas, such that both processes contribute to the total light yield of the prompt signal.
This visible energy $E_{prompt}$ is directly correlated with the incident antineutrino energy $E_{\bar{\nu}_e}$: 
\begin{equation}
E_{prompt}= E_{\bar{\nu}_e}- 0.784\,{\rm MeV}.  
\label{Epro}
\end{equation}
The neutron quickly thermalizes in the proton-reach media and is captured by a proton with the mean capture time
$\tau = 259.7 \pm1.8$\,$\mu$s~\cite{Cosmogenic}. The capture of neutron on a proton is accompanied by the emission of a gamma with 2.22\,MeV energy, that provides a delayed signal.
The space-time coincidence of the prompt and the delayed signals provides a clean signature of the ${\bar \nu_{e}}$ interaction.
To select the ${\bar{\nu}_e}$ candidates we apply the following criteria: 
the coincidences are searched for with $\Delta$t in the range 20--1280\,$\mu$s ($\sim$5 neutron capture time) and a distance between the reconstructed positions $\Delta$r~$<$ 1\,m,  that accounts for the uncertainty of the spatial reconstruction algorithm and the free path of the 2.2\,MeV $\gamma$'s. 
To reduce the external $\gamma$-background from radioactive decays in the detector materials, we accept only candidates having the prompt event position reconstructed  inside the inner detector at distance ($D_{IV,prompt}$) larger than 25\,cm with respect to the the time-varying IV surface: $D_{IV,prompt} > 25\,{\rm cm}$.
The energy of the prompt event is required to be above the value corresponding to the IBD threshold, considering the energy resolution ($N_{p.e.,prompt} > $  408 p.e.), while for the delayed event, the energy cut was tuned to cover the gamma peak from neutron capture on proton ($860 <N_{p.e.,delayed} < 1300$ p.e.). The lower limit is justified because photons at the edge of the scintillator can escape, depositing only a fraction of their total energy.
A pulse shape discrimination cut is applied, requiring the Gatti parameter~\cite{gatti} of the delayed events to be less than 0.015. This cut is effective to remove the time-correlated  $\beta + (\alpha+\gamma)$ decays of $^{214}$Bi-$^{214}$Po, having a time constant close to the neutron capture time.  Such background was relevant only during the detector purifications in 2010-2011 as a consequence of the increased radon contamination. 

Among the cosmogenic isotopes, only $^{8}$He~ ($\tau$ = 171.7\,ms, $Q$ = 10.7\,MeV) and $^{9}$Li~($\tau$ = 257.2\,ms, $Q$ = 13.6\,MeV) have decay modes with electrons and neutrons in the final state, indistinguishable from  the IBD. Since the half-life of these nuclides is of the order of 200\,ms, a time window of 2\,s after a muon crossing the scintillator is chosen in this analysis to effectively remove these events. 
Finally, to  further decrease the neutron-related backgrounds,  we reject all events having a neutron-like event 2\,ms before the prompt or after the prompt or the delayed events.
The present search for ${\bar \nu_{e}}$ fluxes is based on the data acquired between December 2007 and October 2017 during a total live-time  of 2771 days. 
After the application of the selection cuts the total live-time decreases to 2485 days (6.8 years).
The resulting efficiency of all cuts has been estimated by using the Borexino MonteCarlo code to be $\varepsilon$=(85.0 $\pm$ 1.5)\%, and the total exposure for the present data set is 1494 $\pm$ 60 tons per year (100\% efficiency).

\subsection{Backgrounds}
\label{subsec:dsnb_antinu_backg}

\begin{figure}[h]
	\begin{minipage}{1\linewidth}
		\begin{center}
			\includegraphics[width=1\linewidth]{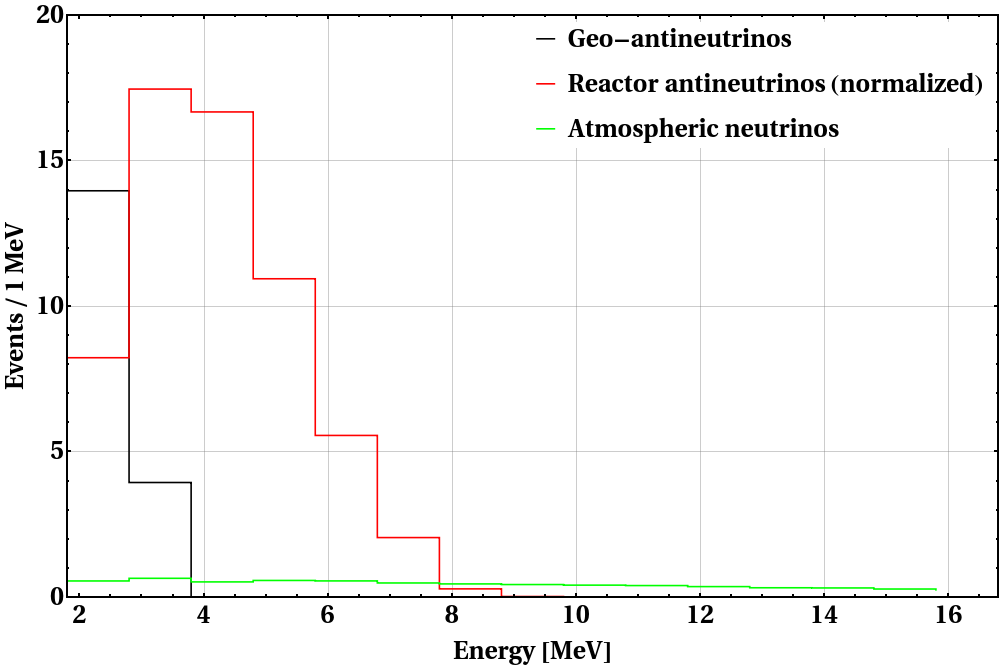}
		\end{center}
	\end{minipage}
	\caption{The number of expected $\nu$-induced background events for each energy bin (black -- geo-${\bar{\nu}_e}$, red -- reactor $\bar{\nu}_e$, green -- atmospheric ${ \nu}$).}
	\label{anti_nu_bck}
\end{figure}

The most relevant sources of $\bar{\nu}_e$ events below 10\,MeV are the Earth's radioactive isotopes and the nuclear reactors, while at higher energies the atmospheric neutrino background dominates the energy spectrum (see Fig.\ref{anti_nu_bck}). In the following, we provide the details about each of these backgrounds.

\subsubsection{Geo-neutrinos}
\label{subsubsection:dsnb_antinu_bckg_geo}

Geo-${\bar \nu_{e}}$ spectrum extends up to 3.26\,MeV, but $^{238}$U and $^{232}$Th chain isotopes are the only energic  enough to significantly contribute events above the IBD  threshold.
The largest contribution to the signal is expected from the rocks closest to the detector, a few hundreds of kilometers around the experimental site.
Detailed models of the crust composition in the Gran Sasso area have been developed, based on geological surveys~\cite{coltorti}. The prediction for the overall signal from $^{238}$U and $^{232}$Th  in the crust (local rocks + rest of the crust) is $S_{geo}$(crust) = 23.4 $\pm$ 2.8\,TNU (1\,TNU= 1\,event/year/10$^{32}$ target protons with 100\% efficiency and for IBD interactions). In the calculation the effect of neutrino oscillations has also been included.

The signal from the mantle is much more uncertain: Earth models provide values for the total U and Th mass in the mantle that span over more than one order of magnitude and the detected signal depends on the distribution of U and Th inside the Earth. 
For all these reasons,  it is possible to construct models perfectly consistent with the geochemical and geophysical constraints but providing very different geo-${\bar{\nu}_e}$  signals, ranging from 0.9 to 33\,TNU.

Borexino has  already made a comprehensive study of  geo-neutrinos ~\cite{BXgeo1,BXgeo2,BXgeo3} but the achieved precision on mantle ${\bar{\nu}_e}$  signal is still poor, $S_{geo}(Mantle) = 20.9^{+15.1}_{-10.3}$\,TNU \cite{BXgeo3}. 
Since we want to quote conservative limits, the minimal expected number of events for each background is considered here. For the present analysis we have therefore chosen the Minimal Radiogenic Earth model, which only includes the radioactivity from the crust and that, in our  case, corresponds to the already mentioned (23.4 $\pm$ 2.8) TNU, i.e. 17.9 $\pm$ 2.1 events in our data sample.

\subsubsection{Reactor antineutrinos}
\label{subsubsection:dsnb_antinu_bckg_rea}

The spectrum of reactor $\bar{\nu}_e$ is more energetic with respect to the one from geo-$\nu$ and is significant till $\sim$10\,MeV.  The $\bar \nu _{e}$  flux comes primarily from
the beta decays of neutron-rich fragments produced in the fission of four isotopes: $^{235}$U, $^{238}$U, $^{239}$Pu, and $^{241}$Pu.
The expected fluxes can be estimated from the knowledge of the monthly energy production at each reactor site,  including the neutrino propagation effects.
At present, there are about 440 nuclear power reactors in the world, providing, nominally, a total amount of about 1200 Thermal GW, corresponding to about 400 Electrical GW. 
With $\sim$200\,MeV average energy released per fission and 6 $\bar{\nu}_e$ produced along the $\beta$-decay chains of the neutron-rich unstable fission products, a reactor with a typical thermal power of 3\,GW emits $10^{20} \,  \bar{\nu}_e$ s$^{-1}$.

\begin{figure}
	\begin{minipage}{1\linewidth}
		\begin{center}
			\includegraphics[angle=0,width=1.03\textwidth]{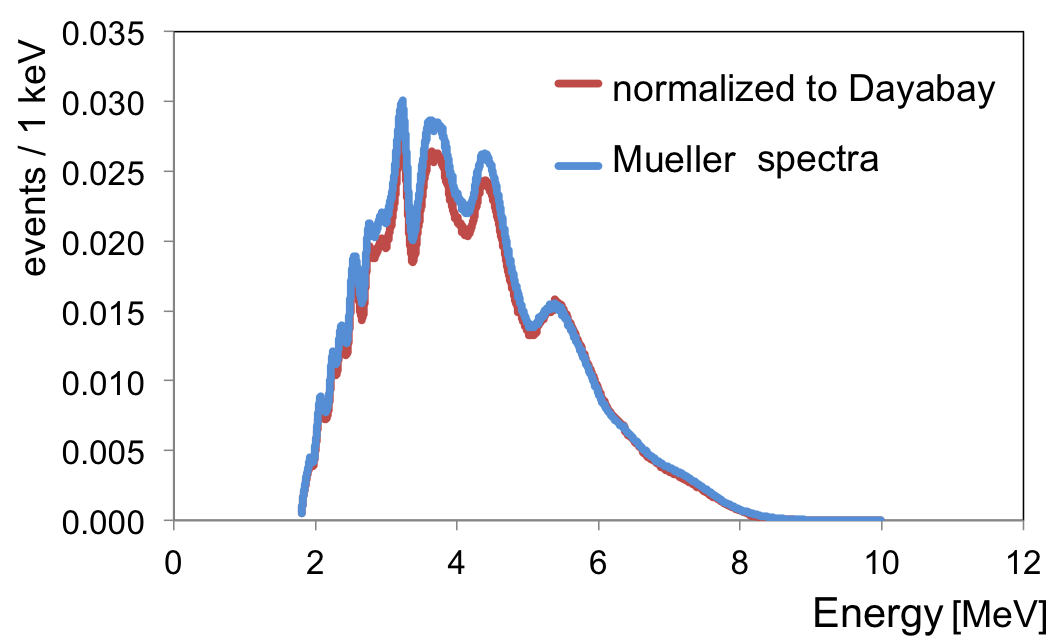}
		\end{center}
	\end{minipage}
\caption{Reactor antineutrino spectra expected in Borexino, assuming 100\% efficiency in the period December 2007 - October 2017, with Mueller spectra (blue) or with the normalization to the Daya Bay measurement (red). }
\label{fig_spettro}
\end{figure}

In the framework of the geo-$\bar{\nu}_e$ study, a precise calculation of the expected signal at the Borexino site has been developed~\cite{Baldoncini2015}.
For each nuclear core, the nominal thermal power and the monthly Load Factors, i.e. the fractions of the nominal power really produced as a function of time are detailed by the International Agency of Atomic Energy (IAEA)~\cite{IAEA}.   IAEA provides the electrical Load Factors, not the thermal ones and the calculation assumes them to be identical. 
For each core, the distance is calculated taking into account the position of the Borexino detector \cite{BXgeo2} (lat = 42.4540 $^{\circ}$N, long = 13.5755 $^{\circ}$W)
and  the positions of all the cores in the world according to the database in  \cite{Baldoncini2015}.  To propagate neutrinos  we use the  three mixing parameters determined by NU-FIT 3.2 (2018) normal hierarchy \cite{Nufit322018}.
The calculation of  $\phi_i(E_{\bar{\nu}_e})$, the $\bar \nu _{e}$ energy spectrum for each fuel component ($i$) at the source, ~deserves particular attention.
The results from reactor antineutrino experiments Daya Bay~\cite{Dayabay2016}, Double Chooz~\cite{abe2014improved}, RENO~\cite{Reno}, and NEOS~\cite{Neos} coherently show that the measured IBD positron energy spectrum deviates significantly from the spectral predictions of Mueller \textit{et al.}~\cite{Mueller2011}. Besides an average deficit of $\sim$6\% in the whole spectrum (the so called "Reactor Anomaly"), a more pronounced peak  between 4--6 MeV is visible in the energy distribution of  the measured events, the so called  "5\,MeV bump."
The ratio of the extracted reactor antineutrino spectrum to prediction deviates from unity, see for instance Fig.3 of ref.\cite{Dayabay2016}.
In order to take into account this effect, we operate in the following way: first we calculate the neutrino spectra corresponding to all four isotopes according to the parameterization of Mueller \textit{et al.}~\cite{Mueller2011}. Then we multiply the total spectrum by an energy-dependent correction factor based on the Daya Bay high precision measurement (extracted from lower panel of Fig.~3 in~\cite{Dayabay2016}).
By following this procedure the same correction factor is applied to all the reactor fuel components ($^{235}$U,$^{238}$U,$^{239}$Pu,$^{241}$Pu). This is an approximation since more likely only a few decay branches have not correctly been accounted for in \cite{Mueller2011}. However, it can be accepted since these branches are not yet precisely known, and we are only interested to judge the possible impact of the reactor spectrum uncertainty on our results.
The calculated antineutrino signals in Borexino, by applying (or not) the normalization to the DayBay measurement, are  shown in Fig. \ref{fig_spettro} as a function of energy.
The corresponding signals  are  of $79.9 ^{+1.4}_{-1.3}  $\,TNU and  $84.8 ^{+1.5}_{-1.4}$\,TNU, respectively, i.e. there is a difference of $\sim$6\%, greater than  the total  uncertainty ($\simeq$2.5\%) on the expected signals, as quoted  in \cite{Baldoncini2015}. We notice that the normalized spectra provide, on average, lower signals because of the 6\% correction and that the "5\,MeV bump," to some extent, compensates this deficit, making the number of events expected more similar at energies above 4.5\,MeV.
Since the normalized spectrum provides the lowest signal, in the computation of the upper limits we have considered this option being the most conservative: this assumption means to attribute 61.1 $\pm$ 1.7 events among our candidates to reactor $\bar{\nu}_e$.

\subsubsection{Atmospheric neutrinos}
\label{subsubsection:dsnb_antinu_bckg_atm}
The most serious background to the detection of DSNB fluxes at energies above 10 MeV is induced by the atmospheric neutrinos, i.e., the $\nu$'s and  $\bar{\nu}$'s generated in the decay of secondary particles  produced in the interactions of primary cosmic rays with Earth atmosphere.

Atmospheric neutrinos may give both Charged Current (CC) and Neutral Current (NC) interactions with the atoms constituting the Borexino scintillator. The most copious isotopes are $^1$H ($6.00\times 10^{31}$/kton), $^{12}$C ($4.46\times 10^{31}$/kton), and $^{13}$C ($5.00\times 10^{29}$/kton). Besides the IBD reaction itself, there are several reactions with $^{12}$C and $^{13}$C nuclei that may, in some cases, mimic the IBD. They have the form of $\nu + A \to \nu(l) + n + \dots + A'$, where $A$ is the target nucleus, $A'$ is the nuclear remnant, $l$ is the charged lepton produced in case of CC processes, $n$ is the neutron, and dots are for other produced particles like nucleons (including additional neutrons) and mesons (mostly $\pi$ and $K$ mesons). The calculation of the induced signal by all these processes is of course quite a complicated task: a dedicated simulation code was therefore developed to precisely quantify this background in Borexino.

Here we summarize the key points:
for energies above 100\,MeV, the atmospheric neutrino fluxes are taken from the HKKM2014 model~\cite{Honda 2015}, while below 100\,MeV the fluxes from the FLUKA code~\cite{Battistoni 2005} have been adopted. The models were chosen as the most up-to-date and precise in respective energy regions. The neutrino fluxes averaged from all directions are considered, since the scintillation light is isotropical and it does not provide  sensitivity to the direction. We calculated the flavor oscillations during neutrino propagation through the Earth, including the matter effects, with the modified Prob3++ software~\cite{ProbPlusPlus} that comprises 1\,km wide constant-density layers according to the PREM Earth's model~\cite{PREM}.
The neutrino interactions with $^{12}$C, $^{13}$C, and $^{1}$H nuclei are generated with the GENIE Neutrino MC code (version 3.0.0, tune G18\_10b)~\cite{GENIE}.
GENIE output final state particles are used as input particles for the G4Bx2 Borexino MC~\cite{g4bx} that allows us to reproduce the detector response. Since the Borexino MC chain results in output files with the same format as real data, the same events filtering code can be applied. 

Taking into account the number of particles in the Borexino scintillator, the resulting interaction rates are 261/year in the total IV plus buffer mass (1181.5 tons). After the simulation of detector response and by applying the same selection cuts as for real data (Sec. \ref{subsec:dsnb_antinu_data}), 9643 IBD-like events were selected out of $2.3 \times 10^6$ interactions. They do correspond to 6.5 IBD-like events in the present analysis statistics of 6.8 years and in the equivalent $\bar{\nu}_e$ energy window 1.8--16.8 MeV.

The uncertainty on this result comes mostly from two sources. Atmospheric fluxes are assumed to be known with $\sim$25\% precision \cite{Battistoni 2005, Honda 2015}. To quantify the uncertainty related to the interaction cross sections we repeated the calculation by using GENIE version 2.12.10 and the rest of the simulation chain, unchanged: the expected number of events decreased by 36\%, probably as a consequence of the cross sections and intranuclear cascade models' differences between GENIE versions. In order to account for other small and unknown uncertainty sources, and assuming that these uncertainties are independent, we consider a conservative uncertainty of 50\%.

In the light of the large uncertainty on this background source, a conservative choice would be to not consider it at all in the upper limit calculations on extraterrestrial $\bar \nu_e$  fluxes. Nonetheless, since it represents the main source of background at energies above $\sim$10\,MeV, we decided to quote both limits obtained with and without atmospheric neutrino background.

\subsubsection{Random coincidences and other non-$\bar \nu_e$ backgrounds} 
\label{subsubsection:dsnb_antinu_bckg_ran}

The space and time correlation of the $\bar \nu _e$ interactions helps to reduce the coincidences rate of non-correlated events. Since they are mainly due to gammas penetrating the detector from outside,  the choice of the fiducial volume (25 cm from the IV) is useful to guarantee an effective suppression. To quantify the fraction of  persisting events, we used an off-time coincidence window of 2--20 s. The number of selected coincidences is then scaled to the 1260 $\mu$s wide time window adopted in the  $\bar \nu_e$ search. A correction factor has to be applied: it takes into account the smaller loss of exposure in the  case of the $\bar{\nu}_e$ candidates search due to the fact that the 2\,s muon vetoing windows before the prompt and the delayed events are partially overlapped while in the case  of the random coincidence search they are not.
The correction factor is about 10\%, and it was estimated by means of a toy MC from the measured muon rate: the total number of expected accidental coincidences after the correction is 0.418 $\pm$ 0.006 in the whole data sample.

Among the other minor non-$\bar{\nu}_e$ backgrounds are ($\alpha$,n) processes: neutrons of energies up to 7.3\,MeV may be emitted by $^{13}$C($\alpha$,n)$^{16}$O reactions following the $^{210}$Po decay, as investigated by KamLAND~\cite{Abe2008}. In Borexino, due to low level of intrinsic $^{210}$Po contamination, this source of background yields as few as $\sim$0.2 events in the present statistics that can be neglected in the upper limit calculation, always by taking a conservative approach.

\subsubsection{Summary of backgrounds}
\label{subsubsection:summary_of_background}

Table \ref{bckg} contains a summary of the relevant background rates.

\begin{table}[h]
	\caption{Estimated numbers of background events among the $\bar{\nu}_e$-candidates. For the reactor signal, we report the events expected if the normalization to Daya Bay measurement is applied, see Sec.\ref{subsubsection:dsnb_antinu_bckg_rea}. The quoted errors are the ones due to the systematical uncertainties.}
	\begin{center}
	\begin{tabularx}{0.4\textwidth}{|c|c|}
		\hline
		Background source & Expected  events\\
		\hline
		Reactor~$\bar{\nu}_e$ &61.1 $\pm$ 1.7  \\
		Geo $\bar{\nu}_e$& 17.9 $\pm$ 2.1\\ 
		Atmospheric neutrinos& 6.5 $\pm$ 3.2 \\
		Accidental coincidences& 0.418 $\pm$ 0.006\\
		\hline
		\hline
		Total: & 85.9 $\pm$ 4.2\\
		\hline
	\end{tabularx}
	\label{bckg}
	\end{center}
\end{table}

\subsection{Selected events}
\label{subsec:events}

With the main DAQ system, 101 ${\bar{\nu}_e}$ candidates have been identified, passing all selection cuts. A cross-check in parallel was then performed  for each candidate by using the data from the FADC system that provides a linear dynamic range to higher energies and an independent input for pulse shape analysis.
The properties of the observed events for each energy bin and the expected backgrounds can be found in Table \ref{table_res_flux}.
\begin{figure}[h]
	\begin{minipage}{1\linewidth}
		\begin{center}
			\includegraphics[width=1\linewidth]{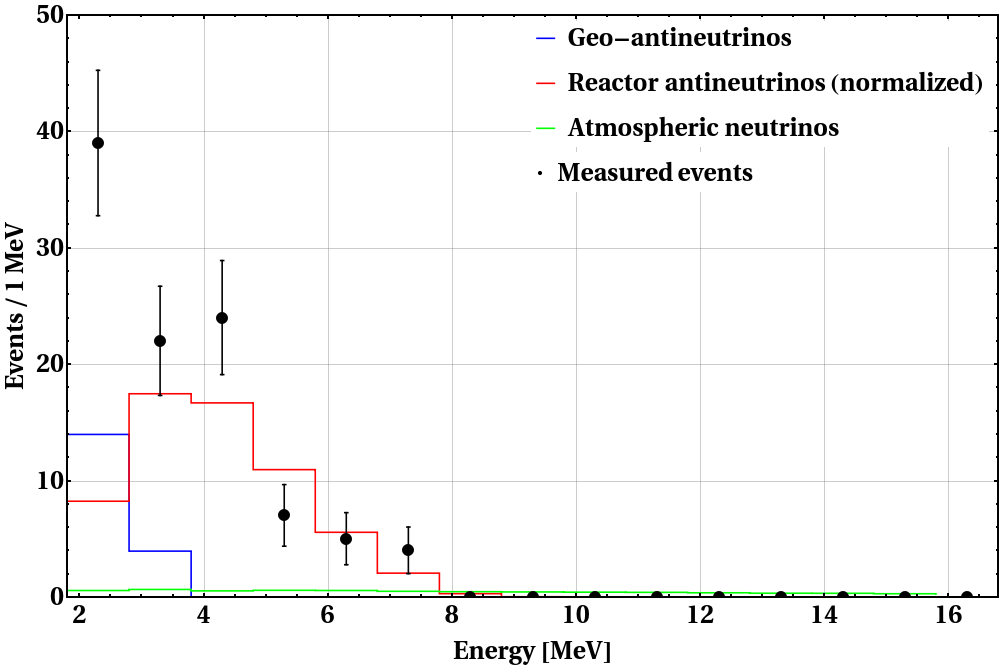}
		\end{center}
	\end{minipage}
	\caption{The number of measured events (black data point) is compared for each energy bin with the expected backgrounds (blue -- geo-${\bar{\nu}_e}$, red -- reactor ${\bar{\nu}_e}$, green -- atmospheric ${ \nu}$) (Sec. \ref{subsec:dsnb_antinu_backg}).}
	\label{anti_nu_bck_meas}
\end{figure}
All the candidates have energies below 7.8\,MeV, and most of them  are concentrated in the very first energy bins (1.8--4.8\,MeV), a feature understandable on the base of the spectral shape of the dominant backgrounds, i.e., geo-$\bar{\nu}_e$ and reactor $\bar{\nu}_e$ (see Fig. \ref{anti_nu_bck_meas}).   
An excess of the measured events with respect to backgrounds is clearly visible in the lowest energy bin (1.8--2.8\,MeV). The reason is that for the geo-${\bar \nu_{e}}$  signal, we have assumed the Minimal Radiogenic Earth's model, while this excess is likely an indication for mantle geo-${\bar \nu_{e}}$. 
Note that above 2.8\,MeV (close to geo-neutrino spectrum endpoint) we observe 62 candidates, while 63 $\pm$ 2 events are expected from the backgrounds, in perfect agreement.

\subsection{Model-independent upper limits}
\label{subsec:dsnb_antinu_womod}

The model-independent limit for electron antineutrino flux ($\Phi_{\bar{\nu}_e}$)  in each energy bin ($i$) is defined by the equation:
\begin{equation}
\Phi_{\bar \nu_{e},i} = \frac{N_{90,i}}{<\sigma> \cdot \varepsilon \cdot N_{p} \cdot T}
\label{eq:flux_lim}
\end{equation}
where $N_{90}$ is the 90\% \text{C.L.} upper limit for the number of antineutrino interactions obtained by following the Feldman-Cousins approach \cite{FeldCous}, $<$$\sigma$$>$ is the mean cross-section of Inverse Beta Decay calculated according to \cite{Strumia} for each energy bin, $\varepsilon$=(0.850 $\pm$ 0.015) is the average detection efficiency, $N_{p}=(1.32 \pm 0.06) \times 10^{31}$ is the  number of protons in the Borexino average fiducial volume mass and $T$ = 2485 days is the total live-time. 

\begin{figure}[h]
	\begin{minipage}{1\linewidth}
		\begin{center}
			\includegraphics[width=1\linewidth]{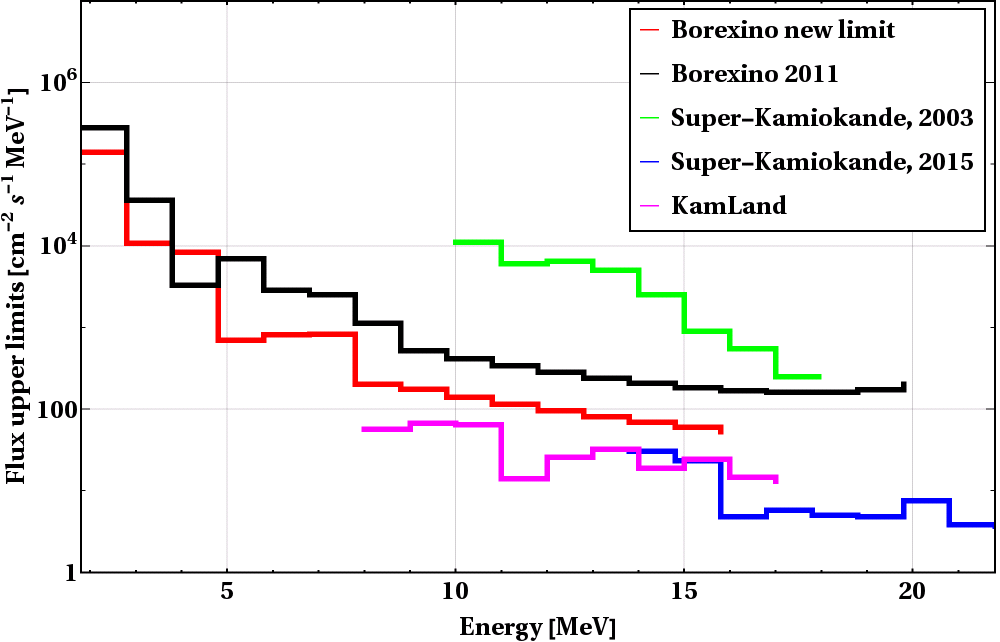}
		\end{center}
	\end{minipage}
	\caption{New Borexino model-independent limits on electron $\bar \nu_e$ fluxes from unknown sources in comparison with the results of other detectors (Super-Kamiokande \cite{Super-K 2003, Super-K 2015}, KamLAND \cite{KamLAND 2012}) and previous Borexino limits \cite{BxAntinu}.}
	\label{limits_anti_plot}
\end{figure}

New Borexino limits are shown in Fig.~\ref{limits_anti_plot}. The other limits existing in literature are quoted on the same plot.
Due to almost five-fold increase in statistics, we improved our previously published limits \cite{BxAntinu} by a factor of 2.5 on average.
Below 8 MeV, Borexino limits are the only existing, thanks to the high energy resolution, the low intrinsic backgrounds, and the small reactor $\bar \nu_e$ flux at the Gran Sasso site.

Due to the large uncertainty of the prediction for the atmospheric $\bar \nu_e$ signal, we show only more conservative limits obtained without this background source. Table~\ref{table_res_flux} contains Borexino results obtained both with and without taking into account atmospheric neutrino background.

\begin{table}[h]
\caption{Energy bin (1 MeV wide bins, the lower energy edge is quoted), number of observed events, number of expected background events without (with) atmospheric neutrino contribution, and 90\%~C.L. upper limits on the $\bar \nu_e$ flux without (with) atmospheric neutrino background. The reactor signal was normalized to the Daya Bay measurement (Sec. \ref{subsubsection:dsnb_antinu_bckg_rea}).}
\begin{center}
\begin{tabular} {|c|c|c|c|}
\hline
 $E\rm{[MeV]}$ & N$_{ev}$ & N$_{bkg}$ & ${\Phi}\rm[{cm^{-2}s^{-1}]}$ \\
\hline

1.8	&39	&22.4 (23.0)	& 1.40 (1.37) $\times~10^{5}$	\\
2.8	&22	&21.5 (22.2)	& 1.07 (1.00) $\times~10^{4}$	\\
3.8	&24	&16.7 (17.2)	& 8.39 (8.13) $\times~10^{3}$	\\
4.8	&7	&10.9 (11.5)	& 6.92 (7.07) $\times~10^{2}$	\\
5.8	&5	&5.55 (6.10)	& 8.08 (7.21) $\times~10^{2}$	\\
6.8	&4	&2.04 (2.52)	& 8.29 (7.68) $\times~10^{2}$	\\
7.8	&0	&0.28 (0.72)	& 2.02 (1.65) $\times~10^{2}$	\\
8.8	&0	&0.01 (0.44)	& 1.75 (1.44) $\times~10^{2}$	\\ 
9.8	&0	&0.00 (0.41)	& 1.40 (1.17) $\times~10^{2}$	\\
10.8	&0	&0.00 (0.39)	& 11.4 (9.59) $\times~10^{1}$	\\
11.8	&0	&0.00 (0.35)	& 9.50 (8.12) $\times~10^{1}$	\\
12.8	&0	&0.00 (0.32)	& 8.05 (7.01) $\times~10^{1}$	\\
13.8	&0	&0.00 (0.31)	& 6.91 (6.03) $\times~10^{1}$	\\
14.8	&0	&0.00 (0.27)	& 6.00 (5.34) $\times~10^{1}$	\\
15.8	&0	&0.00 (0.24)	& 5.27 (4.74) $\times~10^{1}$	\\

\hline
\end{tabular}
\label{table_res_flux}
\end{center}
\end{table}

\subsection{Limits on diffuse supernovae background}
\label{subsec:dsnb_antinu_wmod}
The energy spectra of the observed $\bar{\nu}_e$ events can be compared with the expectations for the different astronomical source models to get hints about the presence of their signal or to quote upper limits on the corresponding fluxes. In general, for the supernova core-collapses, a unique model does not exist.
The basic ideas were established in 1930s~\cite{Zwicky}, but the nature of the shock wave revival and explosion mechanism are still not fully understood. The main problem of the supernovae explosion mechanism theory is an  explanation for the energy transfer of the gravitational energy to the stellar envelope and initiation of the outward shock wave. The mean energy of the emitted neutrinos from Supernova collapse depends on this mechanism and nowadays could only be extracted from numerical simulations.

In current work, we used, as references, the 1d numerical simulations performed by two groups (\cite{Nakazato},\cite{Hudepohl}). These models were chosen because they are  long-term simulations and yield a mean energy value for neutrinos emitted during the collapse.
In the case of the model by Nakazato \textit{et al.}~\cite{Nakazato}, the expected DSNB flux at the Earth, as a function of the neutrino energy, is directly provided as the result of the numerical simulation of SN core-collapse and is made available on a webpage \cite{Nakazato}. For the present study we have selected the predictions for neutrino normal mass hierarchy, including the oscillation effects. In the case of the Hudepohl \textit{et al.} model \cite{Hudepohl}, only the average energies of $\bar{\nu}_e$ emitted during the core-collapse are provided.  We have therefore calculated the DSNB flux on the Earth, starting from Eq. \ref{eq:fluxdsnb}.
The emission neutrino spectrum in our calculation is parameterized as:
\begin{equation}\label{Nsn}
\frac{dN_\nu}{dE_{\nu}} = \frac{(1+\alpha)^{1+\alpha}E_{tot}}{\Gamma(1+\alpha)}\left (\frac{E_{\nu}}{E_{av}}\right)^{\alpha}e^{\left\{-(1+\alpha)\frac{E_{\nu}}{E_{av}}\right\}} 
\end{equation}
where $E_{tot}$ = $3\times 10^{58}$\,MeV (=$5\times 10^{52}$\,erg) is the average neutrino luminosity, $\alpha$ = 4 is a pinching parameter \cite{Keil}, and $E_{av}$ is the average neutrino energy ($E_{av}$ = 11.4\, MeV for $\bar{\nu}_e$ and $E_{av}$ = 9.4\, MeV for $\nu_{e}$). 

The $R_{SN}$ function in Eq. \ref{eq:fluxdsnb} is expressed according to the formula \cite{Rsn}:
\begin{equation}
R_{SN} = \rho_{*}(z) \times \frac{\int_{8}^{50}\psi(M)dM}{\int_{0.1}^{100}M\psi(M)dM} 
\label{eq:lsalp}
\end{equation}
where $\psi(M)dM$ is the number of stars in the mass range $M$ to $M$ + $dM$. According to the Salpeter Initial Mass Function \cite{Salpeter}, the integral ratio is equal to 0.0070 $M_{\odot}$, where $M_{\odot}$ is the mass of the Sun.

For the $\rho_{*}(z)$ expression in Eq. \ref{eq:lsalp} we have taken the broken power-law from \cite{Rsn}:
\begin{equation}
\dot{\rho}_{*}(z)=\dot{\rho}_{0}\left(\left(\frac{z+1}{B}\right)^{\beta  \eta }+\left(\frac{z+1}{\text{C}}\right)^{\gamma  \eta }+(z+1)^{\alpha  \eta }\right)^{1/\eta }
\end{equation}
where $\dot{\rho}_{0}$ = 0.0178~$M_{\odot}$~$\rm{yr}^{-1}Mpc^{-3}$ \cite{Hopkins}, $z$ is the redshift, $\alpha$ = 3.4,  $\beta$ = -0.3, $\gamma$ = -3.5 and B = $(z_{1}+1)^{1-\frac{\alpha }{\beta }}$, C~=~$(z_{2}+1)^{1-\frac{\beta }{\gamma }}(z_{1}+1)^{\frac{\beta -\alpha }{\gamma }}$, $z_{1}=1$, $z_{2}=4$.

\begin{figure}[h]
	\begin{minipage}{1\linewidth}
		\begin{center}
			\includegraphics[width=1\linewidth]{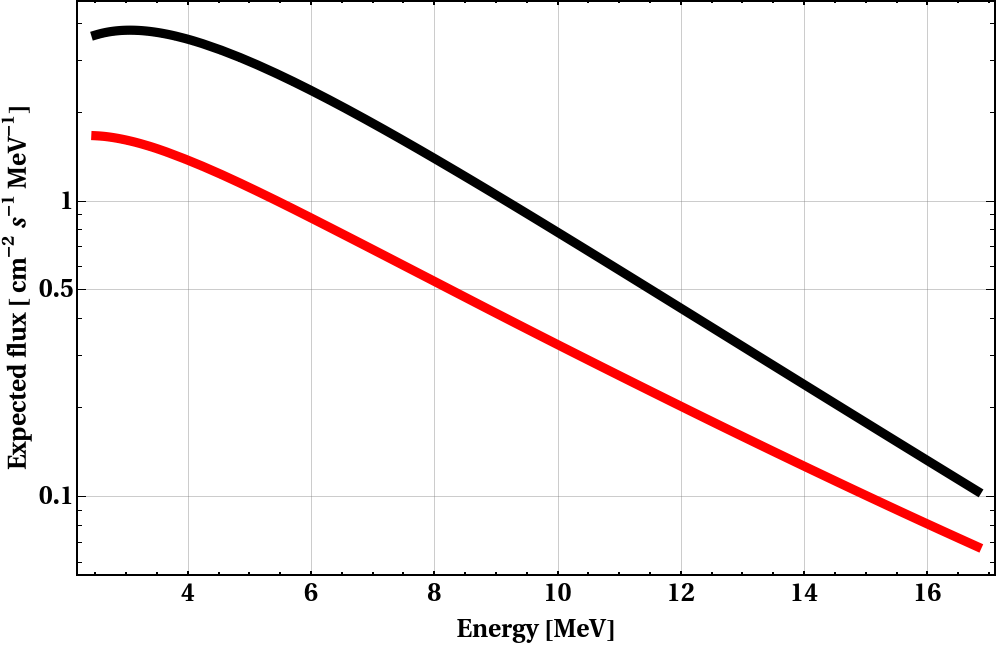}
		\end{center}
	\end{minipage}
	\caption{Expected fluxes for the DSNB $\bar{\nu}_e$ predicted by Nakazato \textit{et al.}~\cite{Nakazato} (red) and Hudepohl \textit{et al.}~\cite{Hudepohl} (black) models.}
	\label{DSNB_sp}
\end{figure}

The expected DSNB fluxes at the detector for both models are reported in Fig.~\ref{DSNB_sp}.
 
On the basis of the energy spectrum of our $\bar{\nu}_e$ candidates, the upper limit at 90\% \text{C.L.} for integral DSNB flux in different energy ranges has been calculated according to:
\begin{center}
	\begin{equation}
	F_{90}=\frac{\int _{E_{min}}^{E_{max}}  {F_{M}(E)} {dE}}{\int _{E_{min}}^{E_{max}}  {N_{M}(E)} {dE} }\times N_{90} 
	\label{eq_90}
	\end{equation}
\end{center}
where $N_{90}$ is the 90\% \text{C.L.} upper limit for the number of $\bar{\nu}_e$ interactions, $F_{M}$ is the flux predicted by a model in the corresponding energy region, and $N_{M}$ is the expected number of events calculated considering the data exposure, the detection efficiency, and the cross section variation over the predicted energy spectrum.

The model-dependent 90\%~C.L. upper limits on the DSNB $\bar{\nu}_e$ flux are presented in Table \ref{table_res_flux_mod}.

\begin{table}[h]
 \caption{Borexino 90\% \text{C.L.} upper limits on the DSNB $\bar{\nu}_e$ flux, according to Nakazato \textit{et al.} \cite{Nakazato} and Hudepohl \textit{et al.} \cite{Hudepohl} models, without (with) the inclusion of the atmospheric neutrino background.}
\begin{center}
\begin{tabular} {|c|c|c|}
\hline
  &{Nakazato \cite{Nakazato}}& {Hudepohl \cite{Hudepohl}} \\
\hline
\hline
    E\rm{[MeV]} &{${\Phi}\rm[{cm^{-2}s^{-1}]}$}& {${\Phi}\rm[{cm^{-2}s^{-1}]}$} 	\\
 \hline
2.8--16.8 & $<$~2.4 (1.7) $\times~10^{3}$ & $<$~2.6 (1.8) $\times~10^{3}$ \\
7.8--16.8 & $<$~106.0 (38.2) & $<$~112.3 (40.5) \\
\hline
\end{tabular}
\label{table_res_flux_mod}
\end{center}
\end{table}

We observe that the choice of the DSNB  model does not affect the result much, while it is very sensitive to the inclusion of the atmospheric neutrino background, with a difference up to factor of 2.7 in the energy range 7.8--16.8\,MeV.
Our more conservative limit for the 7.8--16.8 MeV range, $\Phi_{\bar{\nu}_e}$ $<$ 112.3\,cm$^{-2}$s$^{-1}$ ($90\%$ \text{C.L.}) is slightly better with respect to the one obtained by the KamLAND collaboration \cite{KamLAND 2012} ($\Phi_{\bar{\nu}_e}$ $<$ 139\,cm$^{-2}$s$^{-1}$ ($90\%$ \text{C.L.})) that indeed refers to a larger energy range (8.3--31.8 MeV), and it is to some extent complementary to the one from SuperKamiokande  ($\Phi_{\bar{\nu}_e}$ $<$ 2.9 cm$^{-2}$s$^{-1}$ (90\% \text{C.L.})) for $E_{\bar{\nu}_e} >$ 17.3 MeV \cite{Super-K 2012}.

\subsection{Limits on $\nu_e \rightarrow \bar{\nu}_e$ conversion in the Sun due to spin-flavor precession}
\label{subsec:sfp}

The presence of antineutrinos in original neutrino fluxes can be a consequence of neutrino electromagnetic interactions induced by the non-zero neutrino magnetic moment ($\mu_\nu$) (see \cite{Giunti:2014ixa} for a recent review of neutrino electromagnetic properties). Being roughly proportional to the neutrino mass~\cite{Shrock:1982sc}, $\mu_\nu$ is expected to be non-zero at the light of oscillation paradigm with massive neutrinos. 
Neutrinos with anomalous $\mu_\nu$ interacting with strong magnetic fields in the Sun may undergo spin-flavor precession (SFP), which changes their helicity and, possibly, flavor \cite{PhysRevLett.45.963}. Dirac neutrinos under SFP transit into a sterile right-handed state, while for Majorana neutrinos spin-flip is equivalent to $\nu_\alpha-\bar\nu_\beta$ conversion. Under the CPT conservation, this process for Majorana neutrinos is necessarily accompanied by the flavor change, and thus, the appearance of $\bar\nu_e$ in the Sun can be described as a combined effect of SFP and neutrino oscillations in matter (MSW effect):
\begin{align}
\nu_e \xLongrightarrow{\quad\mathrm{SFP}\quad} \bar\nu_\mu \xLongrightarrow{\quad\mathrm{MSW}\quad} \bar\nu_e \\
\nu_e \xLongrightarrow{\quad\mathrm{MSW}\quad} \nu_\mu \xLongrightarrow{\quad\mathrm{SFP}\quad} \bar\nu_e 
\end{align}

The tightest limit on the $\nu_e-\bar\nu_e$ conversion probability was obtained in the KamLAND experiment and is equal to $5.3\cdot10^{-5}~(90\%~\text{C.L.})$ ~\cite{KamLAND 2012}. In Borexino, a study of $\nu_e-\bar\nu_e$ conversion was previously performed in \cite{BxAntinu} using $\sim$2~years of data acquired during Phase~I. In the following, we update our previous results using a larger dataset.

\subsubsection{Search for antineutrinos from the ${}^{8}\text{B}$ reaction}
\label{B8}

The same IBD candidates as for the previous studies have been used here to search for the $\bar\nu_{e}$ from the conversion of $^8\text{B}$ neutrinos having energies up to 16.8~MeV. We developed separate analyses in two energy regions, 1.8--7.8~MeV (LER), and 7.8--16.8~MeV (HER).

In the HER, we used the Feldman-Cousins approach~\cite{FeldCous} to get the $90\%$~C.L. upper limit ($N_{90}$) on the antineutrino interaction rate for this energy region. Then the limit on the antineutrino flux ($\Phi_{\mathrm{lim}}$) is obtained by following the same approach as in subsection~\ref{subsec:dsnb_antinu_womod}, Eq. \ref{eq:flux_lim}.
The average IBD cross section $<$$\sigma$$>$ considered in this study is weighted over the undistorted spectrum of $^8\mathrm{B}$ neutrinos. 

The analysis in the LER is instead performed by applying the spectral fit procedure developed for geo-neutrino studies~\cite{BXgeo3}. The fit is performed assuming contributions from (1) geo-, reactor and atmospheric neutrinos; (2) $\bar\nu_e$-like background from  accidental coincidences, $(\alpha,n)$ reactions and cosmogenic isotopes; (3) $^8\mathrm{B}$ antineutrino spectrum. Assuming that SFP probability is not energy dependent, the spectral shape of $^8\mathrm{B}$ antineutrino coincides with the neutrino spectrum.

In the HER we expect 0.3 background events in the region of interest, assuming the reactor background spectrum is normalized to Daya Bay measurement ~\cite{Dayabay2016} and the absence of the atmospheric neutrino background. Zero events observed in this region corresponds to $N_{90} = 2.15$, and a limit on the antineutrino flux
\begin{equation}
\label{lim78}
  \phi_{\bar\nu}(E>7.8~\mathrm{MeV})<138.0~\mathrm{cm}^{-2}\mathrm{s}^{-1}.  
\end{equation}
As the region above 7.8~MeV contains $36\%$ of the $^8\mathrm{B}$ flux, the limit in the whole energy range is correspondingly $\phi_{\bar\nu}<383.7~\mathrm{cm}^{-2}\mathrm{s}^{-1}~(90\%~\mathrm{C.L.})$. Taking the Standard Solar Model (SSM) values of $^8\mathrm{B}$ neutrino flux under assumptions of high~(HZ) and  low~(LZ) solar metallicity~\cite{Vinyoles:2016djt}, i.e., the abundance of heavy elements in the Sun, one can obtain limits on the $\nu_e-\bar\nu_e$ conversion probability:
\begin{align}
p^{\mathrm{HZ}}_{ \nu_e \rightarrow \bar{\nu}_e}<7.0\cdot10^{-5}&(90\%~\mathrm{C.L.}),\\
p^{\mathrm{LZ}}_{ \nu_e \rightarrow \bar{\nu}_e}<8.5\cdot10^{-5}&(90\%~\mathrm{C.L.}).
\end{align}
In the LER,  the spectral fit procedure provides an upper limit for the number of events of $N_{90} = 13.3$, by  profiling the $\chi^2$ of the fit result as a function of $^8\mathrm{B}$ antineutrino interaction rate. This limit can be improved by combining both energy regions in the fit, and the final result is $N_{90} = 6.1$, corresponding to 2.19 events above 7.8~MeV and the conversion probability of:
\begin{align}
    p^{\mathrm{HZ}}_{ \nu_e \rightarrow \bar{\nu}_e}<7.2\cdot10^{-5}&(90\%~\mathrm{C.L.}),\\
    p^{\mathrm{LZ}}_{ \nu_e \rightarrow \bar{\nu}_e}<8.7\cdot10^{-5}&(90\%~\mathrm{C.L.}).
\end{align}

The limit calculated with the combined approach appears to be weaker than that obtained from the analysis of the HER only, but we conservatively consider this result as the final one. 
\subsubsection{Neutrino magnetic moment}
\label{nmm}

We have estimated limits on the effective magnetic moment of solar neutrinos, assuming SFP in the solar core and the convective zone, separately.

In the $^8\mathrm{B}$ neutrino production region, conversion probability depends on the transverse component of the strength of the toroidal solar magnetic field. Neutrino magnetic moment can be derived as in ~\cite{Akhmedov:2002mf}:
\begin{equation}
    \mu_\nu \leq 7.4\times10^{-7}\cdot \left( \frac{p(\nu_e-\bar\nu_e)}{\sin^2 2\theta_{12}}\right)^{1/2}\cdot\frac{\mu_B}{B_\perp[\mathrm{kG}]}.
\end{equation}

Taking our final result for the conversion probability under the assumption of high metallicity, namely $p(\nu_e-\bar\nu_e)<7.2\cdot10^{-5}~(90\%~\mathrm{C.L.})$ and $\sin^2\theta_{12} = 0.297$~\cite{Nufit322018}, one can obtain
\begin{equation}
    \mu_\nu < 6.9\cdot10^{-9} B^{-1}_\perp [\mathrm{kG}]\cdot \mu_B~(90\%~\mathrm{C.L.}).
\end{equation}

Due to limited possibilities of magnetic field measurements in the innermost part of the Sun, only marginal values of $B$ in the solar core are provided by solar physics. The most stringent observational limits come from measurements of solar oblateness, which could be distorted by strong magnetic fields in the core~\cite{Friedland:2002is,Antia2008}. According to these investigations, $B<7$~MG. Theoretical studies of the stability of the toroidal magnetic field in the rotating radiative zone~\cite{Kitchatinov2008} provide stronger constraints on the magnetic field: $B<600$~G. These estimations correspond to upper limits of the neutrino magnetic moment between $1.14\cdot10^{-8}\mu_B$ and $1.08\cdot10^{-12}\mu_B$ ($90\%$~C.L.). The latter is stronger than the current limit from astrophysical observations~\cite{Raffelt:1989xu}, while the former is overlaid by other measurements.

SFP can also occur in the convective zone of the Sun via interaction with turbulent magnetic fields~\cite{Miranda:2003yh,Miranda:2004nz,Friedland:2005xh}. Considering the expression for the $\nu_e\rightarrow\bar\nu_e$ conversion probability from~\cite{Friedland:2005xh} and neutrino oscillation parameters from~\cite{Nufit322018}, the neutrino magnetic moment can be expressed as
\begin{equation}
    \mu_\nu \leq 8.0\times10^{-8}\cdot \left(p(\nu_e-\bar\nu_e)\right)^{1/2}\cdot B^{-1}[\mathrm{kG}]\cdot \mu_B.
\end{equation}

As the estimated strength of the magnetic field in the convective zone is of the order of $10^4$~G~\cite{Fan2004}, the corresponding magnetic moment limit is $\mu_{\nu}<3.4\cdot10^{-11}\mu_B$ at 90\% C.L. This value is close to that obtained with the Borexino analysis of solar neutrino data~\cite{agostini2017limiting}. Similar results can be obtained assuming low metallicity SSM.

\subsubsection{Study of the conversion in Low Energy Region}
\label{sfp_low_e}

At energies below the IBD threshold, $\bar \nu_e$  still interacts  via elastic scattering with electrons and thus contribute to the recoil electrons spectrum. Besides the shape distortion, the conversion of neutrino into antineutrino should reduce the detected neutrino rate since the $\bar\nu-e$ cross section is substantially smaller than that for electron neutrinos. Therefore, by constraining solar neutrino fluxes with the SSM prediction~\cite{Vinyoles:2016djt}, one can gain additional sensitivity to the conversion rate. It is worth mentioning that this detection technique is also sensitive to spectral shape distortion due to the electromagnetic $\nu-e$ interaction induced by the non-zero neutrino magnetic moment. This fact was previously used by Borexino in Ref.~\cite{agostini2017limiting} to put a strong bound on $\mu_\nu$ without assumptions on the solar magnetic field.

The previous limit on neutrino-antineutrino conversion ($p_{ \nu_e \rightarrow \bar{\nu}_e}$) obtained by Borexino using the method described in this section is~\cite{BxAntinu}:
\begin{equation}
\label{oldlimit}
p_{ \nu_e \rightarrow \bar{\nu}_e} < 0.35 \: (90\%\text{C.L.})
\end{equation}

In the present work, we improve this limit following the recent progress of Borexino in solar neutrino detection~\cite{Agostini:2018uly}.

In this study we assume MSW as the leading conversion mechanism, and the $\nu \rightarrow \bar{\nu}$ conversion as a sub-dominant process.

The differential cross section of neutrino elastic scattering of electrons for all neutrino flavors 
is given by the expression:
\begin{align}
\frac{d\sigma_{\alpha}(E,T)}{dT} \;=\; \frac{2}{\pi} G_{F}^2 m_e 
\Biggl[ g_{\alpha L}^2 +  \Biggr. \nonumber\\
\Biggl. g_{\alpha R}^2 \left(1 - \frac{T}{E}\right)^2 -
g_{\alpha L} g_{\alpha R} \frac{m_e T}{E^2} 
\Biggr]
\;,
\label{diffCS}
\end{align}
where $G_F$ is the Fermi constant, $m_e$ is the electron mass, and $E$ and $T$ are neutrino and recoil electron kinetic energies, respectively. The coupling constants at tree level are given by expressions:
\begin{align}
\label{gLgRconstants}
g_{\alpha L} & = 
\begin{cases}
\sin^2 \theta_W + \frac{1}{2} & \mbox{for $\alpha=e$,} \\
\sin^2 \theta_W - \frac{1}{2}     & \mbox{for $\alpha=\mu,\tau$,}
\end{cases}
\cr
g_{\alpha R} & = \;\; \sin^2 \theta_W \qquad\quad\mbox{for $\alpha=e,\mu,\tau$.}\vphantom{\Bigg|}
\end{align}

Note that $\nu_{\mu}$ and $\nu_{\tau}$ have the same cross section due to equal coupling constants.  

The differential cross section for antineutrinos has the same form as \eqref{diffCS}, with $g_{L}$ and $g_{R}$ coupling constants swapped. Their values for all three flavors are:
\begin{equation}
\label{antigLgRconstants}
g_{L}  = \;\; \sin^2 \theta_W - \frac{1}{2} \qquad 
g_{R} = \;\; \sin^2 \theta_W.
\end{equation}

Electron neutrinos interact with electrons by both CC and NC, while $\bar{\nu}_e$s interact by NC only and $\nu_e$ has an approximately three times larger cross section than $\bar{\nu}_e$. Moreover, the swap of the coupling constants affects the second and the third energy-dependent terms in \eqref{diffCS} and, therefore, distorts the shape of the spectrum.  

Both $\nu_{\mu/\tau}$ and $\bar{\nu}_{\mu/\tau}$ interact with electrons by NC. Since $ g_{\mu /\tau R}^{2} \approx g_{\mu/\tau L}^{2}$, the shape and normalization of the spectra are almost the same for neutrinos and antineutrinos, and the effect of the shape distortion is less pronounced compared to $\nu_e$'s case. Thus, Borexino is sensitive to $\nu_e \rightarrow \bar{\nu}_e$ conversion only.

Taking into account antineutrino component due to $\nu \rightarrow \bar{\nu}$ conversion, the observed spectra in the detector are given by the expression\footnote{For clarity, we present this formula in energy units, omitting the convolution with the energy response function of the detector}:
\begin{equation}
\label{recoilspec}
\frac{dR_{\nu+\bar\nu}}{dT} 
\;=\; N_{e} \Phi_{\nu} \int dE\, \dfrac{d \lambda_{\nu}}{dE} 
\left[ A_{\nu}(1-p_{ \nu \rightarrow \bar{\nu} }) + A_{\bar{\nu}} p_{ \nu \rightarrow \bar{\nu} }
\right]
\;,
\end{equation}
where $N_e$ is the number of electrons in the fiducial volume, $\Phi_{\nu}$ and $\frac{d \lambda_{\nu}}{dE}$ are total neutrino flux and energy spectrum for a given neutrino producing reaction ($\nu=pp$,
~$^{7}\text{Be}$,~$pep$,~$\text{CNO}$) and $P_{ee}(E)$ is electron neutrino survival probability predicted by MSW-LMA and
\begin{gather}
\label{Anu}
A_{\nu} = \dfrac{d\sigma_{\nu_e}}{dT} P_{ee}(E)
+ \frac{d\sigma_{\nu_{\mu/\tau}}}{dT}
\left(1 - P_{ee}(E)\right), \\
\label{Aantinu}
A_{\bar{\nu}} = \dfrac{d\sigma_{\bar{\nu}}}{dT} . 
\end{gather}

The same event selection criteria and spectral fit procedures developed by the Borexino collaboration for studies of solar neutrinos are used in the data analysis (see ~\cite{Agostini:2018uly,LongPaper,agostini2017limiting}) . The $^7\text{Be}$ contribution was modified to account for the hypothetical $\nu_e\rightarrow\bar{\nu}_e$ conversion. This spectral component provides the most sensitive probe of the possible appearance of the $\bar\nu$ due to the significant $^7\text{Be}$ response changes both in the shape and the amplitude. Contributions from other solar neutrino components are less pronounced.

 Solar neutrino fluxes $\Phi_{\nu}$ are constrained in the analysis with high and low solar metallicity SSM predictions~\cite{Vinyoles:2016djt}. Uncertainties related to the target mass determination and oscillation parameters were found to be small compared to those associated with SSM-flux prediction and are accounted for by using the SSM-penalty terms.  
 
 The fitting procedure consists in maximization of the multivariate likelihood function $L( p_{ \nu_e \rightarrow \bar{\nu}_e },\vec{\theta}\,)$ for a set of $p_{ \nu_e \rightarrow \bar{\nu}_e }$ values. Then the likelihood profile was analyzed to determine the upper bound for $p_{ \nu_e \rightarrow \bar{\nu}_e }$.

The statistics selected for the present study correspond to data acquired from December 14, 2011 until May 21, 2016 (1291.51~days $\times$ 71.3~tons) of fiducial exposure, i.e., the same period already used for the direct study of the magnetic moment of neutrino at detection in Ref.~\cite{agostini2017limiting}. 
The recent advances in understanding of the detector response allowed the fit to be performed in the energy range $0.19\,\mathrm{MeV}<T_e<2.93\,\mathrm{MeV}$, which includes $pp$, ${}^{7}\text{Be}$, $pep$, and CNO electron-recoil spectra \cite{Agostini:2018uly}. 
In this lower energy range, the number of triggered PMT's  ($N_{dt1}$) in the time window 230 ns is preferred as an energy estimator. More detailed information on the analysis procedures and event selection could be found in Ref. \cite{Agostini:2018uly,LongPaper,agostini2017limiting}.

\begin{figure}
 \centering
   \includegraphics[width=0.8\linewidth]{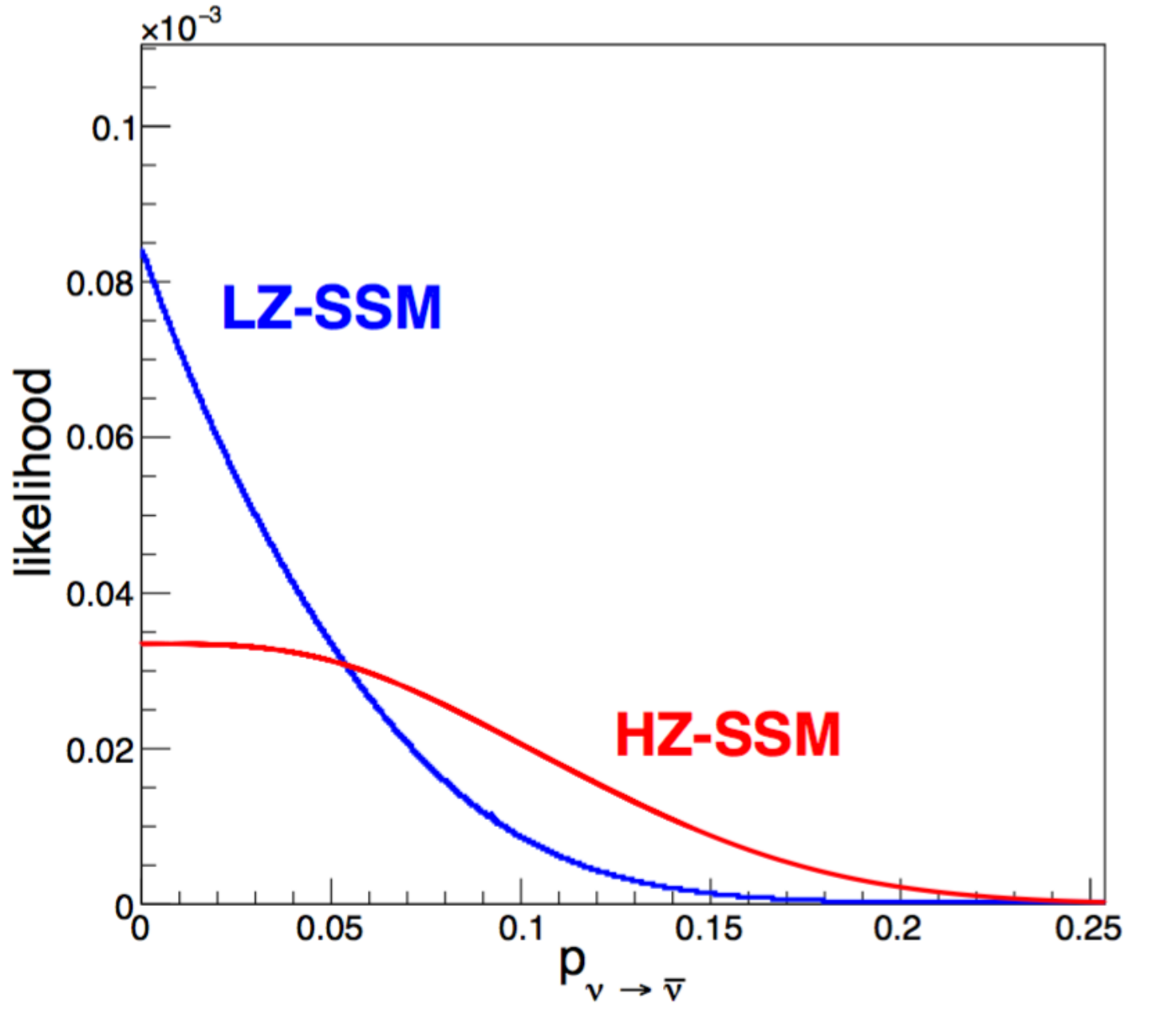}
 \caption{Likelihood profiles for  $\nu \rightarrow \bar{\nu}$ conversion probability obtained with HZ- and LZ-SSM constraints.}
 \label{fig:likelihoodHZLZ}
\end{figure} 

The resulting likelihood profiles for HZ- and LZ-SSM are shown in Fig. \ref{fig:likelihoodHZLZ}.
One can obtain bounds for the  fraction of $p_{ \nu_e \rightarrow \bar{\nu}_e}$ for the HZ-SSM case by numerical integration of the profiles in Fig. \ref{fig:likelihoodHZLZ}:
\begin{equation}
\label{boundHZ}
p_{ \nu_e \rightarrow \bar{\nu}_e}^\mathrm{HZ} < 0.14 \: (90\%\text{C.L.})
\end{equation}
and for LZ-SSM:
\begin{equation}
p_{ \nu_e \rightarrow \bar{\nu}_e}^\mathrm{LZ} < 0.08 \: (90\%\text{C.L.})
\end{equation}

A conservative limit is given by Eq.\eqref{boundHZ}, and it provides an improvement by a factor of two with respect to the limit in Eq.\eqref{oldlimit}, previously obtained in~\cite{BxAntinu}.


\section{Study of events correlated with solar flares}
\label{sec:flares}

Flares are caused by the restructuring of the solar magnetic field, which leads to the acceleration of protons and other charged particles and ions. 
Neutrinos could be generated in the decays of pions, which are abundantly produced in $pp$- and $p\alpha$-collisions in the flare's region.

Neutrino spectrum depends on the spectrum of initially accelerated colliding particles and is poorly known.
Nonetheless, for various sets of input parameters the mean neutrino energy is expected to be around 10\,MeV \cite{Kocharov}.
Production of $\bar{\nu}_e$ is suppressed with respect to $\nu_{e}$ due to a higher threshold of $\pi^{-}$ generation in $pp$-collisions.

The possibility for neutrino emissions correlated with solar flares was first advanced in the eighties by R. Davis \cite{SFlares Homestake-1, SFlares Homestake-2} as an explanation for the excess of events observed in several runs taken by the Homestake Cl-Ar experiment. Homestake run 117 was taken at a time of intense X12 flare (flare class X12 corresponds to the $12 \times 10^{-4}$\,W/m$^{2}$ intensity of the flux in the X-ray band within 1--8\,\AA) on June 4, 1991. Based on the number of observed extra events in that run, an allowed band for the neutrino fluence, compatible with the data, was indicated in \cite{SFlares SNO}.

Here we present a search for $\nu_{x}$ and $\bar{\nu}_x$ ($x=e,\mu,\tau$) from a variety of the solar flares by looking for their elastic scattering on electrons in the Borexino scintillator.

Information about the flares is taken from the GOES database \cite{GOES}.
The database provides the flare's date, the start and end time, and the class.
Assuming the neutrino flux would be proportional to the flare's intensity, we consider the most intense flares of M and X classes, which correspond to intensities of the photon flux higher than 10$^{-5}$\,W/m$^{2}$ and 10$^{-4}$\,W/m$^{2}$, respectively, in the X-ray band within 1--8\,\AA.

The analysis is based on data collected between November 2009 and October 2017. 
A total of 798 flares are selected during this period.
The most intense (X9.3) flare event was registered on September 06, 2017.

The followed approach is to search for an excess of single events above the measured background at the time of a flare.
For the \textit{i}-th flare we choose the time window $\Delta T_{i}^{SIG}$ equal to the flare's duration, according to the database.
The background is calculated in a time window of the same length $\Delta T_{i}^{BKG}$, opened before the $\Delta T_{i}^{SIG}$.
In case the previous flare occurred within this time window, $\Delta T_{i}^{BKG}$ is opened after the $\Delta T_{i}^{SIG}$.
We require at least 95\% of Borexino's up time for both windows: 472 flares of 798 fulfill this criterium, with an overall data coverage in both $ \Sigma \Delta T_{i}^{SIG}$ and $ \Sigma \Delta T_{i}^{BKG}$ of 99.9\%.
The integral intensity over the 472 selected flares is 1.78 $\times~10^{-2}$ \,W/m$^{2}$, i.e., factor of $\sim$15 larger than intensity of the flare occurred during Homestake run 117.

For a first study in the energy range of 1--15 MeV, the data acquired both by the primary and the FADC DAQ systems have been used. Events are selected by vetoing muons and muon daughters both in the $\Delta T_{i}^{SIG}$ and $\Delta T_{i}^{BKG}$ windows: a 2\,s veto is applied after muons crossing ID, and a 2\,ms veto after muons crossing OD.
Single events having energy $E = 1-15$\,MeV are then selected without any fiducial volume cut.

In order to look for neutrino events below 1\,MeV and for a better cross-validation of the whole analysis, an independent study of the data acquired by the primary DAQ alone was performed. This study was done for the same data-taking period and the same $\Delta T_{i}^{SIG}$ and $\Delta T_{i}^{BKG}$ windows, but in the energy range $E = 0.25-15$\,MeV. Events were selected by applying a dynamic fiducial volume cut of 145\,tons (75\,cm from the shape of the IV) to suppress the external $\gamma$ radioactivity. The muon veto duration was chosen to be of only 0.3\,s, because at energies below 3\,MeV, the cosmogenic isotopes are not the dominant background.

\begin{figure}[h]
\begin{center}
\includegraphics[width = 8 cm] {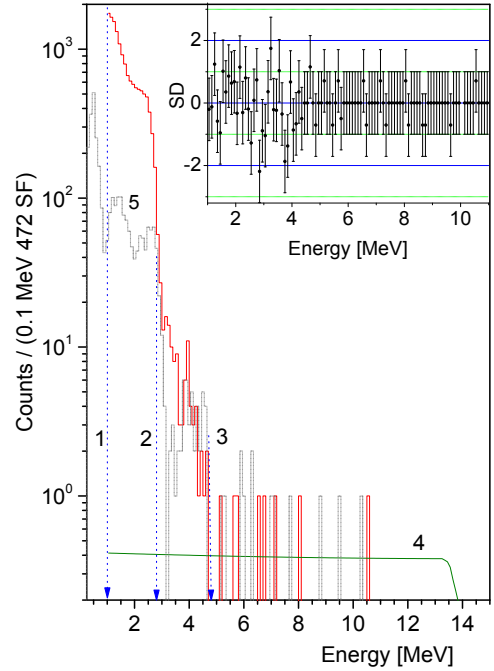}
\caption{Borexino energy spectrum of single events in correlation with solar flares, measured by the FADC system within 1--15\,MeV range (red) and by the primary DAQ system within 0.25--15\,MeV range (grey). In inset, the difference between the spectra measured in $ \Sigma \Delta T_{i}^{SIG}$ and $ \Sigma \Delta T_{i}^{BKG}$ time windows is shown in the units of standard deviations (SD). Blue dotted arrows indicate the three energy regions chosen for the separate analysis (details in text). Line 4 shows the expected spectrum of recoil electrons for 14\,MeV neutrinos per one flare with the fluence $1\times10^{10}~ \rm{cm^{-2}}$.}
\label{Spectra_Koun}
\end{center}
\end{figure}

Figure \ref{Spectra_Koun} shows the energy spectrum of selected single events measured by the FADC system within 1--15\,MeV range (red line) and by the primary DAQ system within 0.25--15\,MeV range (grey), for the integrated time exposure $ \Sigma \Delta T_{i}^{SIG}$.
The difference between this spectrum and the one measured for the $ \Sigma \Delta T_{i}^{BKG}$ is shown in the inset.
No statistically significant excess of events is observed in correlation with the selected flares.

\begin{figure}[h]
\begin{center}
\includegraphics[width = 8 cm] {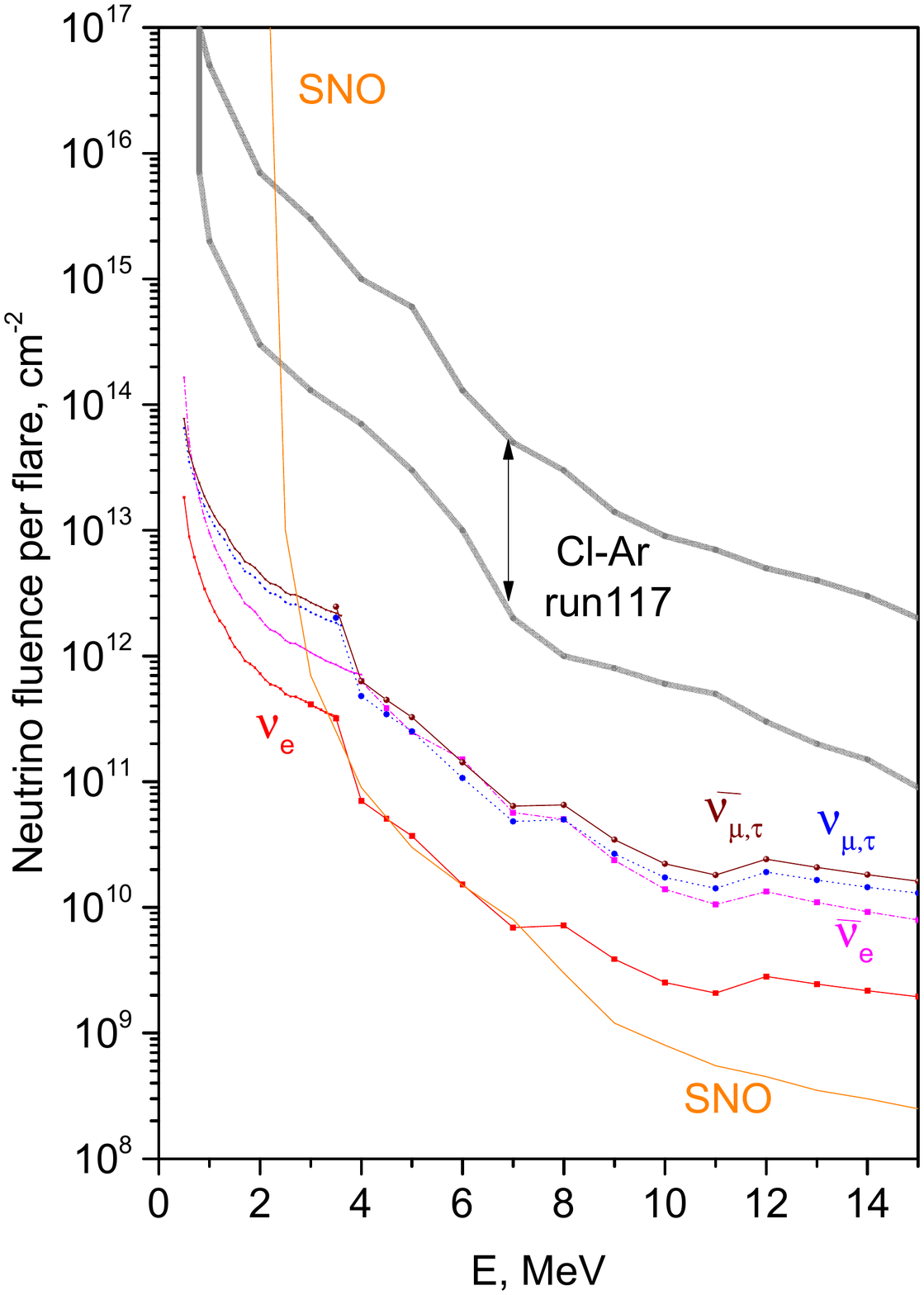}
\caption{Borexino 90\%~C.L. fluence upper limits obtained through neutrino-electron elastic scattering for $\nu_{e}$, $\bar{\nu}_e$, $\nu_{\mu,\tau}$, and $\bar{\nu}_{\mu,\tau}$. In the plot the limits obtained for $\nu_{e}$ by SNO \cite{SFlares SNO} are labelled and the range of fluences that would have explained the Cl-Ar Homestake excess in run 117.}
\label{Limits_472_Flares}
\end{center}
\end{figure}

In order to obtain the fluence upper limits for flare-correlated neutrinos we used the same approach as in \cite{BxGRB}: the fluence limit for neutrinos of energy $E_{\nu}$ is calculated according to the equation:
\begin{equation}\label{eq:ESfluence}
\Phi_{\nu}(E_{\nu})=\frac {N_{90} (E_{\nu}) } {N_e \sigma_{\rm {eff}} (E_{\nu})},
\end{equation}
where $N_e$ is the number of electrons in the Borexino scintillator, that is, $N_e = 9.2 \cdot 10^{31}$ for the whole IV and  $N_e = 4.8 \cdot 10^{31}$ for the 145\,tons FV.
Since the scattering of monoenergetic neutrinos with energy $E_{\nu}$ off electrons leads to recoil electrons with a Compton-like continuous energy spectrum with maximum energy $T_{\nu}^{max} = 2 E_{\nu}^2 / (m_e + 2 E_\nu)$, $\sigma_{\rm{eff}}(E_{\nu})$ is the effective cross section for an interacting neutrino with energy $E_{\nu}$ to recoil electrons with energy $T$ in the interval ($T_{\rm th} > 0,T_{\nu}^{max}$) with 100\% detection efficiency. This effective cross section can be expressed as:
\begin{equation}
\label{eq:ESsigma}
\sigma_{\rm{eff}}(E_{\nu}) = \int _{T_{th}}^{T_{\nu}^{up}} {dT} \int_{T^-}^{T^+}\frac{d\sigma(E_\nu,T')}{dT'} {G(T,T')}{dT'},
\end{equation}
where $T_{\nu}^{up} \cong T_{\nu}^{max}+\sigma_T(T)$.
The Gaussian function $G(T,T')$ with variance $\sigma_T^2(T)$ accounts for the finite energy resolution of the detector, with $T^- = T-3\sigma_T(T)$ and $T^+ = T+3\sigma_T(T)$.
The numerator in Eq.~\ref{eq:ESfluence}, $N_{90} (E_{\nu})$, is the 90\%~C.L. upper limit on the number of flare-associated events per one flare, due to neutrinos with energy $E_{\nu}$, calculated as $N_{90} = (Q(0.9) \times \sqrt{N_{\rm in} + N_{\rm bgr}}) / N_{\rm flares}$, where $Q(0.9) = 1.64$ is a quantile function for normal distribution. 
Here, $N_{\textrm{in}}$ and $N_{\rm bgr}$ denote overall numbers of events in the energy
interval ($T_{th},T_{\nu}^{up}$), detected in the time periods $ \Sigma \Delta T_{i}^{SIG}$ and $ \Sigma \Delta T_{i}^{BKG}$, respectively.

The procedure was repeated for neutrino energies $E_{\nu}$ from 0.5 to 3.5\,MeV in increments of 0.1\,MeV,  from 3.5 to 5\,MeV in increments of 0.5\,MeV, and for $E_{\nu} >$ 5\,MeV in 1.0\,MeV steps.
In order to maximize the signal to background ratio, the lower integration limits $T_{th}$ from Eq.~\ref{eq:ESsigma} was optimized for different neutrino energies considering the shape of the spectrum decreasing with energy (Fig.~\ref{Spectra_Koun}).
The three blue lines indicated in Fig.~\ref{Spectra_Koun} with indices 1, 2, and 3 show the thresholds of electron energies $T_{\rm th}$, for which $E_{\nu}$ was set to (1.5, 3.5)\,MeV, (4.0, 5.0)\,MeV, and (6.0, 15.0)\,MeV intervals, respectively.

In order to set the fluence limits for flare-correlated neutrinos (antineutrinos) of electron and $(\mu + \tau)$ flavors individually, the corresponding cross section in Eq.~\ref{eq:ESsigma} was set to $\sigma_{\nu_e}$ ($\sigma_{\bar{\nu}_e}$) and $\sigma_{\nu_{\mu,\tau}}$ ($\sigma_{\bar{\nu}_{\mu,\tau}}$), respectively.
The results obtained for two DAQ systems independently were found to be consistent within statistical uncertainty of our measurements.

Figure \ref{Limits_472_Flares} and Table \ref{table_fluences} show Borexino limits obtained from the primary DAQ ($E_{\nu}<$ 3.5\,MeV) and the FADC DAQ ($E_{\nu}>$ 3.5\,MeV). 
Limits for $\nu_{e}$ obtained by SNO \cite{SFlares SNO} and an allowed band for the neutrino fluence that would have explained the Homestake run 117 excess of events are also shown for comparison.

As of today, Borexino sets the strongest limits on fluences of all neutrino flavors from the solar flares below 3--7\,MeV. Under assumption that neutrino flux is proportional to the flare's intensity, Borexino's data excludes an intense solar flare occurred during run 117 of the Cl-Ar Homestake experiment as a possible source of the observed excess of events.

\section{Conclusions}
\label{sec:concl}

In this paper we investigated the possible anti-neutrino fluxes from diffuse astrophysical sources such as relic supernovae or the conversion of solar neutrinos into anti-neutrinos in the magnetic field of the Sun. 
The extreme radiopurity of the Borexino detector allowed us to set new limits on diffuse supernova neutrino background for $\bar{\nu}_e$ in the previously unexplored energy region below 8\,MeV, and to get, even with very conservative assumptions, competitive results between 7.8 and 16.8\,MeV.
The new search for $\bar{\nu}_e$ appearance in solar neutrino fluxes was performed both in the energy range $1.8<E_{\bar{\nu}_e}<16.8$~MeV and $0.9<E_{\bar{\nu}_e}<3.3$~MeV. Thanks to an almost 5-fold increase in statistics, we improved previous Borexino limits on neutrino-to-antineutrino conversion ($p_{ \nu \rightarrow \bar{\nu}}$) by a factor of two.
A model-independent study was also presented.

Finally, the most stringent up-to-date limits on fluences for all neutrino flavors from solar flares below 3--7\,MeV have been set. An intense solar flare was excluded as a possible source of the observed excess of events in run 117 of the Cl-Ar Homestake experiment.

\section*{Acknowledgments}
The Borexino program is made possible by funding from INFN (Italy), NSF (USA), BMBF, DFG, HGF, and MPG (Germany), RFBR (16-29-13014, 17-02-00305, 18-32-20073, 19-02-00097), RSF (17-12-01009) (Russia), and NCN (Grant No. UMO 2017/26/M/ST 2/00915) (Poland). 

We also acknowledge the computing services of the Bologna INFN-CNAF data centre
and LNGS Computing and Network Service (Italy), of ACK Cyfronet AGH Cracow (Poland), and of HybriLIT (Russia). 
We acknowledge the hospitality and support of the Laboratori Nazionali del Gran Sasso (Italy).

\begin{table*}[h]
\caption{Borexino 90\%~C.L. upper limits on neutrino fluences from the solar flares.}
\begin{center}
\begin{tabular}{|c|c|c|c|c|}
\hline
$E_{\nu}\rm{[MeV]}$ & ${\Phi}_{\nu_e}\rm[{cm^{-2}]}$ & ${\Phi}_{\bar{\nu}_{e}}\rm[{cm^{-2}]}$ & ${\Phi}_{\nu_{\mu,\tau}}\rm[{cm^{-2}]}$ & ${\Phi}_{\bar{\nu}_{\mu,\tau}}\rm[{cm^{-2}]}$\\
\hline
0.5	&1.83 $\times~10^{13}$	&6.52 $\times~10^{13}$	&1.65 $\times~10^{14}$	&7.73 $\times~10^{13}$ \\
0.7	&6.14 $\times~10^{12}$	&2.57 $\times~10^{13}$	&2.79 $\times~10^{13}$	&3.07 $\times~10^{13}$ \\
1	&2.75 $\times~10^{12}$	&1.29 $\times~10^{13}$	&9.50 $\times~10^{12}$	&1.54 $\times~10^{13}$ \\
2	&7.25 $\times~10^{11}$	&3.81 $\times~10^{12}$	&2.00 $\times~10^{12}$	&4.56 $\times~10^{12}$ \\
3	&4.10 $\times~10^{11}$	&2.23 $\times~10^{12}$	&1.07 $\times~10^{12}$	&2.67 $\times~10^{12}$ \\
4	&2.81 $\times~10^{11}$	&1.56 $\times~10^{12}$	&7.10 $\times~10^{11}$	&1.86 $\times~10^{12}$ \\
6	&1.85 $\times~10^{11}$	&1.04 $\times~10^{12}$	&4.54 $\times~10^{11}$	&1.25 $\times~10^{12}$ \\ 
8	&1.32 $\times~10^{11}$	&7.54 $\times~10^{11}$	&3.21 $\times~10^{11}$	&9.01 $\times~10^{11}$ \\
10	&1.04 $\times~10^{11}$	&5.96 $\times~10^{11}$	&2.51 $\times~10^{11}$	&7.12 $\times~10^{11}$ \\
12	&8.61 $\times~10^{10}$	&4.95 $\times~10^{11}$	&2.07 $\times~10^{11}$	&5.91 $\times~10^{11}$ \\
14	&7.33 $\times~10^{10}$	&4.22 $\times~10^{11}$	&1.75 $\times~10^{11}$	&5.04 $\times~10^{11}$ \\
\hline
\end{tabular}
\end{center}
\label{table_fluences}
\end{table*}

\end{document}